\documentclass[preprint]{aastex}

\begin{document}



\def\gsim{\;\lower.6ex\hbox{$\sim$}\kern-7.75pt\raise.65ex\hbox{$>$}\;}
\def\lsim{\;\lower.6ex\hbox{$\sim$}\kern-7.75pt\raise.65ex\hbox{$<$}\;}
\def\aa{A\&A }
\def\mv{$m_{F555W}$}
\def\mi{$m_{F814W}$}
\def\mj{$m_{F110W}$}
\def\mh{$m_{F160W}$}
\def\mvi{$m_{F555W}-m_{F814W}$}
\def\mjh{$m_{F110W}-m_{F160W}$}
\def\mvh{$m_{F555W}-m_{F160W}$}
\newcommand{\MSUN}{$M_{\odot}$}
\newcommand{\Myr}{M$_{\odot}$~yr$^{-1}$}
\newcommand{\Myk}{M$_{\odot}$~yr$^{-1}$~kpc$^{-2}$}

\title{The Resolved Stellar Populations in NGC~1705
\footnote{Based on observations with the NASA/ESA Hubble
Space Telescope, obtained at the Space Telescope Science Institute,
which is operated by AURA for NASA under contract NAS5-26555}} 

\author{M. Tosi$^1$, E. Sabbi$^2$, M. Bellazzini$^1$, A. Aloisi$^3$, 
L. Greggio$^{1,4}$, Claus Leitherer$^3$, P. Montegriffo$^1$}

\affil{$^1$ Osservatorio Astronomico di Bologna, Via Ranzani 1,
       I-40127 Bologna, Italy\\
       e-mail: bellazzini@bo.astro.it, greggio@bo.astro.it, 
               montegriffo@bo.astro.it, tosi@bo.astro.it}

\affil{$^2$ Dipartimento di Astronomia, Universit\`a di Bologna, Via Ranzani 1,
       I-40127 Bologna, Italy\\
       e-mail: s\_sabbi@bo.astro.it}

\affil{$^3$ Space Telescope Science Institute, 
       3700 San Martin Drive, Baltimore, MD 21218\\
       e-mail: aloisi@stsci.edu, leitherer@stsci.edu}

\affil{$^4$ Universitaets Sternwarte Muenchen, 
       Scheinerstrasse 1, D-81679 Muenchen, Germany\\
       e-mail: greggio@usm.uni-muenchen.de}

\begin{abstract}

We present HST photometry of the resolved stellar population in the dwarf 
irregular galaxy NGC~1705. The galaxy has been observed with both WFPC2 
and NICMOS, and successful images have been obtained in the F555W, F814W, 
F110W and F160W bands. The optical fields cover most of the galaxy, while 
the infrared field (NIC2) maps only its central regions.

The optical photometry provides $\sim$20000 objects down to \mv$\lsim$ 29
in the PC field of view and $\sim$20000 in the three WFCs. In the infrared 
we have been able to resolve $\sim$2400 stars down to \mj, \mh~$\approx$~26. 
A subsample of 1834 stars have been unambiguously measured in all the four 
bands. The corresponding color-magnitude diagrams (CMDs) confirm the existence
of an age gradient, showing that NGC~1705 hosts both young (a few Myr old) 
and very old (up to 15 Gyr old) stars, with the former strongly concentrated 
toward the galactic center and the latter present everywhere, but much more 
easily visible in the external regions.

The tip of the red giant branch (TRGB) is clearly visible both in the
optical and in the infrared CMDs and allows us to derive the galaxy distance. 
Taking into account the uncertainties related to both the photometry and
the TRGB magnitude -- distance relation, we find that the distance modulus 
of NGC~1705 is $(m-M)_0=28.54 \pm 0.26$, corresponding to a distance 
$D=5.1 \pm 0.6$ Mpc.

\end{abstract}


\newpage 
\section{Introduction}

Dwarf irregular and blue compact (BCD) galaxies are extremely useful to 
study galaxy evolution. The closest stellar systems of this type allow 
us to examine in detail phenomena -- like the occurrence of star formation 
(SF) bursts and galactic winds triggered by supernova (SN) explosions, 
the chemical enrichment of the interstellar (ISM) and intergalactic (IGM) 
medium -- which are important not only to understand their cosmological 
evolution, but also to infer the physics directly related to these processes.

Thanks to their low metallicity, late-type galaxies are fundamental to 
infer the primordial $^4$He abundance (e.g. Izotov, Thuan, \& Lipovetsky 
1997; Peimbert, Peimbert, \& Ruiz 2000), and to
check the self-consistency of standard Big Bang nucleosynthesis theories.
This, however, requires not only a careful analysis of the abundances
derived from HII-region spectra, but also a safe understanding of the
relative enrichment of the involved elements ($^4$He, $^{14}$N and $^{16}$O),
which depends both on the SF history and on the gas flows 
in and out of the system.
Galactic winds can be an important factor in galaxy evolution, and are
suggested to be particularly effective in low-mass galaxies with strong SF 
activity, due to the combination of high energy input from SNe and low
potential well.
The detection of gaseous halos around starburst galaxies supports this 
idea (Heckman et al. 1993). The observed chemical properties of irregulars 
and BCDs, including the empirical $\frac{\Delta Y}{\Delta (O/H)}$ relation, 
further suggest that the galactic winds are more enriched in 
elements like oxygen produced by massive stars and ejected by 
Type II SNe (e.g. Matteucci \& Tosi 1985; Pilyugin 1993; Marconi, Matteucci, 
\& Tosi 1994; but see also Larsen, Sommer-Larsen, \& Pagel 2000),
than in elements like helium, mostly produced by lower-mass, longer-lived 
stars. 

The SF history of dwarf galaxies is also crucial to check whether or not
they can represent the local counterpart of the faint blue galaxies found
in excess in deep imaging surveys, as suggested by e.g. Lilly et al. (1995), 
and Babul \& Ferguson (1996). A detailed analysis of the SF history 
in NGC~1569 over the last 0.2 Gyr (Greggio et al. 1998, hereafter G98)
has indeed confirmed that at least some of the most active dwarf galaxies 
have a SF rate (SFR) that would make them contributing to those counts if
substained in the right time interval. What remains to be ascertained is 
the fraction of dwarf galaxies with a sufficiently high SFR, and the existence 
of such strong bursts at epochs corresponding to the redshifts where the excess 
is found (z$\,\simeq\,$0.7--1.0). In this sense it becomes of primary importance 
to check if all the irregulars have formed stars already several Gyrs ago, and 
how many of them have had an intense SF activity in their remote past. As a 
matter of fact, the circumstance that even I~Zw~18, so far considered as the 
prototype of really young galaxies, contains stars born at least several 
hundreds Myr ago (Aloisi et al. 1999, hereafter ATG; Ostlin 2000) is not 
sufficient {\it per se} to demonstrate that none of the dwarfs has started 
only recently ($\lsim$ 40 Myr ago, see Izotov \& Thuan 1999) to form stars.

These questions need to be quantitatively addressed in order to achieve a 
consistent picture on the evolution of late-type galaxies. The nearby 
irregular galaxy NGC~1705 is an ideal benchmark from this point of view. 
It is in fact one of the late-type dwarfs with the best 
documented proof of an ongoing galactic outflow (Meurer et al. 1992, hereafter 
MFDC; Heckman et al. 2001). The outflow is likely to be strongly enriched in 
metals by supernova explosions and stellar winds (Sahu \& Blades 1997), as 
often invoked by theoretical chemical evolution models of dwarf irregulars 
and BCDs (e.g. Marconi et al. 1994).

NGC~1705 ($\alpha_{2000}\,=\,04~ 54~ 15.2,~\delta_{2000}\,=\,-53~ 21~ 40$, 
$l$\,=\,261.08 and $b$\,=\,--38.74) is a dwarf galaxy with apparent magnitudes 
$B$\,=\,12.8 and $V$\,=\,12.3.
Its nucleus hosts a luminous super star cluster (SSC) with estimated mass 
$\sim 10^5$ M$_{\odot}$, probably a proto-globular cluster only 10 Myr old 
(Melnick et al. 1985; O'Connell et al. 1994; Ho \& Filippenko 1996). 
The galaxy has been classified by MFDC as a BCD with a fairly continuous 
SF regime and an approximate oxygen abundance 12+log(O/H)$\,\simeq\,$8.46, 
similar to that of the Large Magellanic Cloud. A more precise oxygen 
abundance of 8.36 has been derived by Storchi-Bergmann, Calzetti \& Kinney 
(1994) from UV, optical and near infrared spectra.
Ground-based work (e.g. Quillen et al. 1995) revealed 
the presence of a composite stellar population, with the older 
($\sim$\,1--10\,Gyr) field population defining the galaxy morphology. 
UV spectra acquired with HST (Heckman \& Leitherer 1997) showed that nearly 
half of the optical/ultraviolet light is contributed by the young stellar 
population, and possibly by the central SSC which may also power the observed 
bipolar outflow. From the lack of spectral stellar wind features, 
Heckman \& Leitherer suggested that the stars in the most luminous SSC of 
NGC~1705 are not more massive than 10--30 \MSUN.

The distance to NGC~1705 is still relatively uncertain, and has been derived 
from CMDs based on pre-Costar HST images ($5\pm2$ Mpc according to O'Connell 
et al. 1995), or from the heliocentric velocity of 628 km s$^{-1}$ 
(4.7 Mpc according to MFDC, 6.2 Mpc according to Meurer et al. 1998).
As shown by Sternberg (1998), at the quoted distances the SSC is
overluminous for its mass and may imply an Initial Mass Function (IMF)
unusually skewed in favor of massive stars.

To study in further detail the stellar populations of NGC~1705 and to infer 
its SF history, we have observed the galaxy with the HST cameras WFPC2 and 
NIC2 in the F555W, F814W, F110W and F160W filters, deriving the CMDs of its 
resolved stars from these datasets. In this paper we present these 
new data and the resulting CMDs (Sections 2, 3 and 4). From the magnitude 
of the TRGB (Section 5) we derive a new estimate of the distance modulus, 
while the comparison of the empirical CMDs with theoretical stellar evolution 
tracks (Section 6) indicates the major evolutionary features of the stellar 
populations in NGC~1705. The results are discussed and summarized in Section 7.

\section{Observations and data reduction}

The infrared observations were performed on 1998 March 17, and repeated 
on 1998 September 19 due to a suspend mode event of NICMOS for a high-energy 
particle hit during the first observing run. We used the F110W and F160W broad 
band filters and the NIC2 camera centered on the brightest SSC, which is near 
the center of NGC~1705.

The optical observations were performed on 1999 March 4--7 with the 
PC camera centered on the SSC and including the NIC2 field. A failure 
in the fine lock guiding mode, coupled with a failure in the star guide 
reacquisition, prevented us from obtaining some of the requested images 
in the four F380W, F439W, F555W, and F814W filters. Only the long exposures 
in F555W and F814W were successfully acquired. The requested long 
observations at shorter wavelength, as well as the short exposures 
in all the four filters, were rescheduled and reacquired on 2000 
November 10--11. The photometric reduction of the bluer filters will 
be presented in a forthcoming paper (Monelli et al. in preparation).

\subsection {The optical data}

The data successfully acquired with WFPC2 in the first observing run 
correspond to eight 1300 s exposures in each F555W and F814W filters. 
The pointing was organized to follow a box dither pattern with CR-split 
0.5. This translates into 4 different pointing positions with a point 
spacing of 0$\farcs$559 for WFPC2, and the following offsets in arcsec: 
(0,0) for Point 1, (0.5,0.25) for Point 2, (0.75,0.75) for Point 3, and 
(0.25,0.5) for Point 4. Two exposures at each dither position were also 
requested for an easier cosmic-ray removal. The dithering technique is 
applied to improve the background estimate, identify hot pixels and 
smooth local pixel to pixel variations from images taken at different 
dither points, and improve the image sampling. A problem in the 
execution of the dither pattern due to a bug in the TRANS software for 
the translation from the Phase 2 commands to the telescope operations,
resulted in 5 exposures at Point 1, and one exposure at each of the other 
Points. Despite this ground-based software failure, the resulting
dither pattern allowed us to correctly remove cosmic rays and better 
sample the PSF, not affecting at all our scientific goal of measuring  
stellar magnitudes with a photometric error $\leq$0.2 mag down to the 
expected limit of $\sim$28 for both \mv~ and \mi. 

For each filter the 8 dithered frames, calibrated through the standard 
STScI pipeline procedure, were combined in a single, fully sampled 
10400 s image, using the software package {\it Drizzle} (Fruchter 
\& Hook 1998). The effective pixel size in the resampled images 
corresponds to 0$\farcs$023 and 0$\farcs$05 for the PC and WFCs 
respectively. The PSF in the resulting {\it drizzled} images has a 
FWHM of 4.2, 3.3, 3.3, and 3.3 pixels in F555W for PC, WFC2, WFC3, 
and WFC4 respectively. The corresponding quantities in F814W are instead 
4.4, 3.3, 3.2, and 3.2.


Fig.~\ref{mosaico_bw} shows a mosaic of the WFPC2 images in the F814W filter, 
obtained with the VmImRegrid task from the ESO VIMOS Data Reduction Software.
The images were regridded to the 
equatorial coordinate system via a gnomonic projection given by the 
WCS information in the image headers. This coordinate system has an 
uncertainty of possibly more than 3$\farcs$5, as indicated by the 
comparison of the WCS values taken from images at different wavelengths. 
The image shows that most of the galaxy falls well within the observed 
fields. The bright spot at the center of the PC field is the SSC, whereas 
the isophotal contour levels divide the galaxy
in regions of different surface brightness, corresponding to different
crowding conditions. Different regions are numbered from 0 (the outermost one) 
to 7 (the innermost region, encircled by the white contour). The indicated
isophotal contours have average flux density in F814W
of 2, 4, 8, 16, 32, 128, 256 counts/px, respectively. 
Throughout this paper we will refer to these zones as
Region 0 ... 7. It is interesting to note that the isophotal contours are
fairly elliptical, smooth (when not disturbed by especially bright objects) 
and concentric.
The only exception is Region 7, which is both off-centered and distorted.
The distortion is due to a sort of spiral pearl necklace (better visible in
Fig.~\ref{pc_nicmos_bw2}) apparently originating from the SSC and leading
north. The SSC is not seated at the center of the isophotal contours,
a circumstance confirming the suggestion by Heckman \& Leitherer (1997)
and Hensler et al. (1998) that the SSC is not at the galaxy center. 


The photometric reduction of the images has been performed with the 
DAOPHOT package in the IRAF\footnote{IRAF is distributed by the National 
Optical Astronomy Observatories, which are operated by AURA, Inc., under 
cooperative agreement with the National Science Foundation.}
environment. We have followed the same procedure described in ATG,
the only difference in this case being that we have detected the stars
independently in the two bands, without forcing in the shallowest frame 
the finding process of the objects detected in the deepest one.
For each detector the objects found in both bands 
were cross-identified with DAOMATCH and DAOMASTER. Only the objects 
with a coordinate offset less than a matching radius of 2.0 pixels
(corresponding approximately to half of the FWHM of the PSFs) were 
eventually retained. After this match we remained with 19838 objects 
having a measured magnitude in both F555W and F814W filters in the PC field. 
The corresponding numbers are 8525 for the WF2, 5711 for the WF3, and 5593 
for the WF4.

The instrumental magnitude of each object in each filter and detector was 
estimated via a {\it PSF-fitting technique}. A PSF template was constructed 
by considering the most isolated and clean stars, sparsely distributed over
each frame: 4 stars were considered for the PC, 5 for the WF2, 4 for the WF3, 
and 6 for the WF4. The zero point of the PSF photometry was determined with 
the standard {\it aperture photometry technique} within a 2 pixel radius 
aperture.
The conversion of the instrumental magnitudes $m_{\rm i}$ (with the exposure 
time already taken into account) to the HST VEGAMAG system was performed by 
following the prescriptions in Holtzman et al. (1995a,b), with the 
updated values provided by the STScI web page (see also Biretta et al. 2000):
\par\noindent
\centerline { $m = m_{\rm i} + C_{\rm ap} + C_{\infty} + ZP_{\rm V} + C_{CTE} $.}
\par\noindent
$C_{\rm ap}$ is the classical aperture correction to convert the photometry 
from a 2 pixel to a conventional 0$\farcs$5 radius aperture for the WFPC2 
(22 and 10 pixels for PC and WFs, respectively, in our {\it drizzled} images).
It has been directly calculated by considering the encircled energy of 
single stars in our observational frames. The second aperture correction 
$C_{\infty}$ is an offset of --0.10 (irrespective of filter and detector) to 
convert the magnitude from the 
0$\farcs$5 radius into a nominal infinite aperture. The zero points  
$ZP_{\rm V}$ are taken from Baggett et al. (1997). We 
have also considered the {\it charge transfer efficiency} (CTE) correction 
$C_{CTE}$ in its new formulation by Whitmore, Heyer, \& Casertano (1999).  
Other secondary calibrations have been neglected. The {\it contamination} 
correction with the new rates derived from Baggett \& Gonzaga (1998) is 
practically zero for all the detectors in the two 
F555W and F814W filters. We have instead not applied any correction 
for the {\it short vs long anomaly} (see e.g. Casertano \& Mutchler 1998), 
because its value has recently turned out to be more uncertain than 
previously thought (Casertano, 2000 private communication).


For a safe interpretation of the characteristics of the galaxy stellar 
populations, we need to distinguish as much as possible {\it bona fide} 
single stars from extended, blended or spurious objects. We have thus 
applied to our catalogs selection criteria based on the shape of the 
objects. To this aim we have considered the Daophot parameters $\chi^2$ 
and {\it sharpness}: $\chi^2$ gives the ratio of the observed pixel-to-pixel 
scatter in the fit residuals to the expected scatter calculated from a 
predictive model based on the measured detector features, while {\it sharpness} 
sets the intrinsic angular size of the objects. We have selected only the 
objects with $\chi^2\leq\,$4 in both filters for the PC, $\chi^2\leq\,$2.2 in 
F555W and $\chi^2\leq$\,2.5 in F814W for the three WFs. Moreover only detections 
with --\,0.5$\,\leq\,sharpness\,\leq\,$0.5 in all filters and for all detectors 
have been eventually retained. These $\chi^2$ and {\it sharpness} values turned
out to be those allowing to best reject spurious and extended objects, without
eliminating also the bright stars.
By inspecting the rejected objects, we have recognized several candidate
star clusters (i.e. fairly round but extended objects) and background 
galaxies. We have 17 star clusters (besides the brightest SSC) in the PC,
4 in the WF2, 5 in the WF3 and 3 in the WF4. The candidate background 
galaxies are 13 in the WF2, 11 in the WF3, and 18 in the WF4. In the 
latter case, the 18 galaxies occupy only a restricted zone of the WF4 and
the concentration is much higher than in the other fields: we are 
presumably dealing with a real, yet unclassified, 
galaxy cluster, since the derived galaxy density (18 arcmin$^{-2}$ down 
to \mi=22) is almost a factor of 5 higher than that predicted by deep 
counts of field galaxies (e.g. Pozzetti et al. 1998).

A further selection often applied is based on the photometric error 
$\sigma_{DAO}$. The distribution of $\sigma_{DAO}$ in the F555W and F814W
bands is shown in Fig.~\ref{dao_sig} for both the PC and the three WFs. 
If we apply the $\sigma_{DAO}$ selection criterion to the stars which 
have already been constrained in the $\chi^2$ and {\it sharpness} 
parameters, with $\sigma_{DAO}\leq\,$0.2 in both filters we retain 
15299 stars in the PC, and 6785, 2812 and 4279 in the WF2, WF3 and WF4, 
respectively. Going down to $\sigma_{DAO}\leq\,$0.1 implies 9224 stars 
retained in the PC, and 4691, 1551 and 2376 in the WF2, WF3 and WF4, 
respectively.

We have put all the measured stars on the common reference frame shown in the
mosaic image of Fig.~\ref{mosaico_bw}, and we have divided them into six 
groups according to their position with respect to the isophotal contour 
levels. Regions 1 and 2, and Regions 4 and 5, are considered together because 
both their stellar populations and their crowding 
conditions turned out to be very similar to each other.
Region 7, despite the low number of stars, has been considered
separately from the others, because it contains the brightest SSC and 
crowding dramatically affects its photometric accuracy and completeness, 
as will be shown from the artificial stars experiments (see Section 4).

\subsection{The infrared data}

For each filter we have acquired ten 512 s exposures with a spiral 
dither pattern of 0$\farcs$2 for a total integration time of about 85 min. 
The dither technique was chosen to account for pixel-to-pixel 
non-uniformities, and to better sample the PSF, which is a very important 
component in the treatment of crowded stellar fields such as that of NGC~1705. 
The MULTIACCUM readout mode was adopted for each exposure in order to get rid 
of saturated pixels and cosmic ray events. To this purpose 
we considered the pre--defined sequence STEP256, which consists of 13 
non-destructive array reads of 0.0 s (the bias level), 0.3 s, 0.4 s, 
1 s, 2 s, 4 s, 8 s, 16 s, 32 s, 64 s, 128 s, 256 s and 512 s. The resulting 
combined F110W image is shown in Fig.~\ref{pc_nicmos_bw2} together with an 
enlargement of a PC portion taken from Fig.~\ref{mosaico_bw}. The white 
contours delineate Regions 6 and 7 as defined in the previous sub-section.

The data reduction has been performed following the procedure described 
by Aloisi et al. (2001, hereafter Al01) for the twin set of NIC2 data acquired 
for the galaxy NGC~1569, for the same HST proposal (GO 7881) and with the 
same technique. The automatic CALNICA pipeline in the IRAF STSDAS package 
was applied to remove the instrumental signatures and co-add datasets from
multiple iterations of the same exposure. Particular attention was paid  
to the necessary corrections for all the instrumental anomalies, artifacts 
and instabilities which are not automatically taken into account 
by the standard pipeline. Among these are the {\it shading}, the {\it pedestal}, 
the {\it cosmic ray persistence}, and the {\it linearity}, all effects that 
we have already considered in Al01. Other anomalies (see Dickinson 
1999 for details on all the possible anomalies) were found in the data 
reduction of NGC~1705 NICMOS images. {\it Grot} and  the 
{\it photometrically challenged 
column} are bad imaging regions, that we had to mask in our data. {\it Bands} 
and {\it bars} are image defects directly related to bias problems. 
Our data were affected by both of them: the use of NIC1 in parallel mode 
originated bands, while the circumstance that we didn't use NIC3 at all 
induced bars in the NIC2 images. We were able to effectively remove both 
bands and 
bars by using the most updated NICMOS-related software available in April 2000. 
Last but not least, our images were also affected by {\it electronic ghosts}, 
also known as {\it Mr Staypuft anomaly}, false faint images due to the 
effects of overexposures (related in 
our case to the brightest SSC of NGC~1705). We got rid of this image anomaly 
by using an algorithm kindly made available by Luis E. Bergeron of the STScI 
NICMOS team.

For each filter the dithered CALNICA calibrated exposures were combined into 
a single frame with the same {\it Drizzle} package used for the optical data (see 
previous subsection). The resulting {\it drizzled} image, with a pixel size of
0$\farcs$0375, has a FWHM of 3.0 pixels in F110W and 3.9 pixels in F160W, instead 
of the original 1.4 and 1.8 pixels in F110W and F160W, respectively, of the 
no-resampled single frames.

The photometric analysis was performed with the IRAF package DAOPHOT. 4142 objects 
in the F110W frame, and 3321 in the F160W frame, were independently detected by the 
routine DAOFIND at 1.5 $\sigma$ above the local background. The instrumental magnitude 
of each object was estimated via PSF fitting with a zero point calculated within a 2 
pixel radius aperture. We adopted as PSF templates the 4 stars of the field which 
turned out to be more isolated and unaffected by profile distortions.

The instrumental magnitudes from the PSF-fitting technique were calibrated in 
the HST VEGAMAG system with the formula given by Dickinson (1999):
\par\noindent
\centerline { $m = m_{\rm i} + C_{\rm ap} + C_{\infty} + ZP_{\rm V}$}
\par\noindent
where $C_{\rm ap}$ is the aperture correction to convert the magnitudes
within a 2 pixel radius in the {\it drizzled} images into the magnitudes within
the conventional 0$\farcs$5 radius (in our {\it drizzled} case it corresponds to 
13.33 pixels and not to the standard 7.6 pixels), $C_\infty$\,=\,--0.152 mag is the 
quantity given by Dickinson to convert the magnitude at 0$\farcs$5 into the
nominal infinite aperture, and $ZP_{\rm V}$ is the published zero point for the 
HST VEGAMAG system (22.381 for F110W and 21.750 for F160W).

The objects found in both bands were cross-identified with the routines
DAOMATCH and DAOMASTER, retaining only those identifications with a difference 
in their coordinates within a radius of 2.0 pixels (roughly half of the PSF 
FWHM in F160W). In this way we selected 2373 objects having a measured magnitude
in both F110W and F160W.

As for the WFPC2 objects, 
to distinguish as much as possible {\it bona fide} 
single stars from extended, blended or spurious objects, we have applied 
to our catalog the $\chi^2$ and {\it sharpness} criteria.
After a visual screening of
each individual object which could or could not be removed depending on
the adopted limits for acceptable {\it sharpness} and $\chi^2$, we have
retained all the objects with  --0.3\,$\,\leq$\,{\it sharpness}\,$\,\leq\,$0.3.
In this case a restriction in $\chi^2$ 
would have inevitably removed also bright {\it bona fide}  single stars.
Of these selected objects, 1707 stars have $\sigma_{DAO}\leq\,$0.2 in both 
filters (see Fig.~\ref{dao_sig}) and 829 stars have $\sigma_{DAO}\leq\,$0.1. 

By inspecting the rejected objects, we have recognized all the candidate 
star clusters found in the portion of the PC image falling in the NIC2
field. They are marked in 
Fig.~\ref{pc_nicmos_bw2} as open circles. If we assume that the galactic
barycenter is most likely located near the center of Region 6 and of all
the other isophotal contours, except Region 7, it is interesting to note
that the candidate  clusters are not evenly distributed around the center
and seem to avoid the area close to the SSC. We consider this as a real
segregation since we have not found any reason to ascribe it to selection
effects in the candidate cluster identification. The properties of the candidate
clusters will be discussed in a forthcoming paper (Monelli et al. in 
preparation) when the
F380W and F439W images will also be examined and the colors will be measured.

As already done for the optical data, the stars have been divided into 
groups, using the same isophotal contours of the WFPC2 F814W mosaic image. 
Due to the smaller size of NIC2, these groups contain stars only from 
Regions 4-5, 6 and 7. Region 7 includes the most luminous SSC and 
most of the brightest stars.

We have also cross-identified the stars appearing in both the PC and 
NIC2 field of view by considering a matching radius of 2 pixels (about half 
of the PSF FWHM in all the four filters). Our final sample includes 1834 
stars and 11 candidate clusters with measured magnitudes in all the four 
F555W, F814W, F110W and F160W frames.

\section{Color-magnitude diagrams}

Fig.~\ref{cm_vi} shows the \mi~{\it vs}~\mvi ~CMDs of all the stars measured
in the seven Regions described in the previous section. The number of 
objects in each panel is labelled in its lower right corner. As mentioned
above, Regions 1 and 2 are plotted together because both their populations
and their crowding conditions are very similar. The same
occurs for Regions 3 and 4.
The spatial distributions of the stellar populations in the various Regions
are clearly different from one another. Bright stars are mostly concentrated 
toward the galactic center and only a few of them are present in 
the outer regions. Instead, faint stars are probably everywhere but 
are more easily resolvable in the external regions, thanks to the less 
crowded conditions. 

The two innermost Regions present a well defined blue and red plume. The
blue plume is located at \mvi$\,\simeq\,$--\,0.1, with the brightest stars at 
\mi=20.3, and corresponds to the main-sequence (MS) evolutionary phase and to 
the hot edge of the core helium burning phase. The red plume is at 
\mvi$\simeq$1.7 and extends up to \mi=19.5. It is populated by red supergiants
(RSGs) at the brighter magnitudes and asymptotic giant branch (AGB) stars at 
fainter luminosities. The
brightest stars, of whatever color, are all located in Region 7, 
and correspond to the pearl necklace visible in Fig.~\ref{pc_nicmos_bw2}
and described in the previous Section. The severe crowding of Region 7
prevents a reliable detection of
objects fainter that about 25.5 in either the F555W or the F814W band.
Region 6 is sufficiently less crowded to allow for the detection
of much fainter objects, and already shows the concentration of stars 
with 1$\,\leq\,$\mvi$\,\leq\,$1.8 and 24.2$\,\leq\,$\mi$\,\leq\,$26,
which is much better delineated from Region 5 outwards and corresponds
to low--mass, old, stars in the red giant branch (RGB) evolutionary phase.
Regions from 3 to 0 contain very few, if any, stars outside the RGB.

Fig.~\ref{cm_vjh} shows in the upper panels the infrared CMDs of the galaxy 
field covered by NIC2 (Regions 4--5, 6 and 7). 
In this case, only a few stars are found in the bluest part 
of the diagram. This is not due to a real lack of young stars (as we know 
from their conspicuous presence in the optical CMD), but to the intrinsic 
faintness of hot stars in the infrared bands. The red plume at \mjh\,=\,1 is 
instead very well sampled and shows a tight vertical finger with \mh~ from 
18 to 20 plus a clump of fainter stars with 0.6$\,\leq$\,\mjh\,$\leq\,$1.6.
Once again, the brightest stars are concentrated in Regions 6 and 7, with
the bluer objects visible only in the latter one.

The three lower panels of Fig.~\ref{cm_vjh} show the CMDs with the largest 
color baseline for the objects common to all the four bands: the different 
evolutionary phases are better separated, but the overall distribution 
does not change. The CMD in other bands of the same objects can be seen in
Annibali et al. (2001).
In order to check the fraction of spatially coinciding objects in the 
four filters with physically meaningful colors and therefore correctly 
cross-identified, we have inspected
the position of the 1834 stars in all the possible combinations 
of CMDs in the four bands. We find that the objects keep consistently their 
colors in the various band combinations (e.g. blue and red objects in 
\mvi~ remain blue and red, respectively, also in the other colors).
This circumstance indicates that we are dealing with intrinsic, physical 
features of real stars and not with spurious colors due to chance spatial 
coincidence of different objects. 
Having already removed from this sample the objects turned out to be 
candidate clusters or galaxies on the basis of the shape selection criteria, 
we can consider these 1834 objects as the safest sub-sample of {\it bona 
fide} single stars in the central region of NGC~1705.

Notice that this positive result is not always achievable in such
crowded fields, both because of the high probability of finding multiple 
objects spatially overlapping and because of the very different fluxes 
of cool and hot stars in different photometric bands. As discussed in
detail by Al01,
the probability of success depends on the combination of crowding, 
wavelength separation of the available bands, and SF history of the examined
region. For instance, in the case of NGC~1569, for which we had no F814W image
available, we have been able to unambiguously cross-identify only a few
stars present in the F110W and F160W frames (analyzed by Al01) and in the F439W 
and F555W frames (analyzed by G98). This is because NGC~1569
is very crowded, and has experienced a strong SF burst in the last 
0.1 Gyr and a conspicuous SF activity between 1.5 and 0.15 Gyr 
ago. The recent burst provides a large number of hot stars, bright in blue 
filters but faint in red ones, while the previous SF episode provides a 
large number of cool stars, bright in red filters but faint in blue ones. 
These two populations are spatially coincident and it is thus inevitable 
that the flux peaks detected in different passbands actually correspond 
to different stars.
The wavelength selection effect is also what causes the blue plume of the 
optical CMD to be poorly defined in the near-infrared (NIR) diagram of
Fig.~\ref{cm_vjh} 
and, conversely, the vertical red finger of the NIR CMD to be sparser in
the optical one. This effect has been found and discussed also in other
galaxies (see, e.g., Schulte-Ladbeck et al. 1999b for VII Zw 403).
The artificial stars experiments described in the next section
give a more quantitative description of these wavelength selection effects.

\section{Role of incompleteness and blending}

The photometry at the faint magnitudes reached by our data and in such
crowding conditions is certainly affected by incompleteness and blending. 
In order to quantify these factors as a function of
magnitude, we have performed a series of artificial stars experiments,
following the procedure briefly described here.

\subsection{Artificial stars experiments}

Artificial star experiments have to probe the observational effects
associated with the whole process of data reduction of a given frame, i.e., for
instance, the accuracy of the photometric measures, the crowding conditions, the
ability of the PSF-fitting code in estimating the sky level or in resolving
partially overlapped sources, etc. It is of the utmost importance that the
artificial stars do not interfere with each other since in that case the output
of the experiments would be biased by {\em artificial} crowding, not present in
the original frame. To avoid this potentially serious bias we 
have divided the frames in grids of cells of known width, and 
we have randomly positioned {\em only one artificial star
per cell} at each run (a similar procedure has recently been adopted by
Piotto \& Zoccali 1999). The additional constraint is that each star must
have a distance from the cell edges sufficiently large to guarantee that all 
its flux and background measuring regions fall within the cell.
In this way we can control the minimum distance between adjacent stars. At each 
run the absolute position of the grid is randomly changed in a way that, after 
a large number of experiments, the stars are uniformly distributed in 
coordinates. 

The stars were distributed in magnitude with a function similar to the observed
luminosity function (LF), except for an excess of faint stars below the 
detection limit of our observations. This was to probe with sufficient statistics 
the (faint) range of magnitudes where the incompleteness is expected to be most 
severe. We have simulated about $10^5$ stars for each WFPC2 camera and filter, 
and about $2\,\times\,10^5$ for each NIC2 image. The whole series of performed
experiments provides $1.2\,\times\,10^6$ artificial stars, for which we have
memorized input and output magnitudes, and any other useful parameter.
Stars with input--output magnitude $\Delta m > 0.75$ were considered 
{\it lost} because such a difference implies that they fell on a real star of 
their same luminosity or brighter.

The artificial stars have been projected onto the same 
reference frame of Fig.~\ref{mosaico_bw} and separated in the corresponding 
radial regions (six annuli for the F555W and F814W filters, and three for the 
F110W and F160W filters) to allow the characterization of incompleteness and
blending in the different regions of the images. The frames have been
re-reduced following exactly the same procedure as for the real stars, and
the same selection criteria for $\chi^2$ and {\it sharpness}
(see Sections 2.1 and 2.2) have been applied.

Fig.~\ref{dm_v} and Fig.~\ref{dm_i} show the input--output magnitude of the
artificial stars resulting from our tests. $\Delta$\mv~ and $\Delta$\mi~ 
are plotted as a function of \mv~and \mi, respectively, for each 
Region from 0 to 7. Fig.~\ref{dm_jh} shows instead
$\Delta$\mj~and $\Delta$\mh~as a function of the input
\mj~and \mh~for the three inner 
Regions. The solid lines superimposed to the plotted distributions report the mean
$\Delta m $ (central lines) and the $\pm 1 \sigma_m$ around the mean. The
value of $\sigma_m$ at three reference magnitude levels are also
reported in each panel to provide an easier and quantitative comparison.

Each of the figures presented above provide a complete and statistically robust
characterization of the photometric errors and of the effects of blending 
affecting our observations. In what follows we will refer to Fig.~\ref{dm_v} as
an exemple to describe the general trends that are present in all these plots.

\begin{itemize}

\item The $\sigma_m$ provide the best estimate of the random component of
the actual photometric error
affecting our data. By comparing them with the $\sigma_{DAO}$ plotted in
Fig.~\ref{dao_sig}, it is apparent that the latter increasingly underestimate 
the actual errors towards fainter magnitudes.

\item The mean $\Delta m $ provides a full description of the systematic
component of the photometric error, due to the average effect of
blending as a function of magnitude. In all panels the mean $\Delta m $
begins to deviate from zero around $m_{F555W}\sim 26$, becoming increasingly 
large
toward faint magnitudes. This means that many stars are recovered with a
brighter magnitude than their input one, because they are blended with a
fainter star that is present in the original frame. Approaching the 
limiting magnitude of the observations (at $m_{F555W}> 28$) artificial stars are
recovered {\em only} if they are pushed up above the detection threshold by 
blending with an undetected real star and/or a positive fluctuation of the
background noise. 

\end{itemize}

We will make full use of the extensive set of artificial star experiments  for 
the production of proper synthetic CMDs that will allow the reconstruction of 
the star formation history 
of the galaxy. This part of the analysis will be described in a
companion paper (Annibali et al. in preparation). 
In the following, we shortly comment on the
effects of blending and incompleteness to assess the overall accuracy of our
photometry, with particular attention to the impact on the distance estimate
discussed in Section 5.

\subsection{Blending and completeness}

From the inspection of Fig.~\ref{dm_v}, Fig.~\ref{dm_i} and Fig.~\ref{dm_jh} 
it is clear that the very concentrated population of bright stars 
in Region 7 produces a high degree of incompleteness at faint 
magnitudes in this region.  The high degree of crowding and the high
level of sky background produced by the wings of the many bright stars results
in much larger photometric errors and brighter limiting magnitude
than in all the other Regions (0-6). 
Outside of this very central part, the behavior of $\Delta m $ as a function of
magnitude is quite similar and very well sampled everywhere. 

It is particularly interesting to check at which magnitude the effects of blending 
become significant, i.e. at what level $\Delta m $ significantly deviates from zero.
First of all we note that the mean value of $\Delta m $ is always lower than 
$\sigma_m$, thus random errors dominate over the systematics associated with the 
blending at every magnitude. Second, in the Regions from 0 to 6, the mean $\Delta m $
is lower than 0.05 mag  for $m_{F555W}<24$, $m_{F814W}<23$, $m_{F110W}<22.5$ and 
$m_{F160W}<21.5$ respectively. Therefore the effect of blending can be considered 
negligible down to very faint magnitudes. The above limits become significantly 
fainter if the most external regions are considered (0-2). In Region 0, which covers 
a large fraction of all the WF cameras, the average $\Delta m $ is still virtually 
null at $m_{F555W}=26$ and $m_{F814W}=25$ (see also Fig.~\ref{cm_corr_vi}, below).

The {\em completeness factor} $CF(m)=N_{out}/N_{in}$ (where $N_{in}$ is the 
number of artificial stars added and $N_{out}$ is the number of artificial
stars successfully recovered for a given magnitude bin in a given passband $m$)
is a measure of the probability of detection of  a real star at a given 
magnitude, as long as blending is negligible. When blending is not negligible
a correct evaluation of $CF(m)$ should take it into account.
In a CMD, the global completeness factor is given by the product of the 
factors in the two involved passbands, hence it is also a function of the color 
index.
In Fig.~\ref{cm_corr_vi} four curves representing levels of equal 
$CF(m_{F814W},m_{F555W}-m_{F814W})$ are superimposed to the CMDs of the
different regions in which we have divided our sample. From top to bottom they 
correspond to $CF=0.95$ (thick line), $CF= 0.75$ (thin line),
$CF= 0.50$ (thick line), and $CF= 0.25$ (thin line). Note that in the
outer Regions (0-5) the $CF=0.50$ level occurs at $m_{F814W}>24$ for any 
given color in the observed range. In particular, the TRGB
(at $m_{F814W}\sim 24.5$ and $m_{F555W}-m_{F814W}\sim 1.6$) lies  
$\sim 1.5$ mag above this level.

An evenly spaced grid of blending vectors is also superimposed to the
CMDs (thin lines with black dots) in Fig.~\ref{cm_corr_vi}. 
The amplitude of these vectors is 
the average $\Delta m$ described above, the dots indicate the
direction of the vectors. The effect of such vectors become appreciable only for
$m_{F814W}\ge 24$ in Regions 5 and 6, and $m_{F814W}\ge 26$ for Regions
0-4 and only at extreme colors.

The analogous plot for the infrared and infrared/optical CMDs is presented in 
Fig.~\ref{cm_corr_jhv}. In the outer Regions (4-6) the $CF=0.50$ level occurs 
at $m_{F160W}\sim 22-22.5$ in both the CMDs and the effect of blending becomes
significant over the whole observed color range only for $m_{F160W}\ge 23.5$.

\section{A new estimate of the distance}

While the use of the TRGB luminosity as a
standard candle dates back to 1930 (see Madore \& Freedman 1998, and references
therein), the development of the method as a safe and viable technique is
relatively recent (Lee, Freedman, \& Madore 1993). In the last few years
it has become a widely adopted technique (see, for instance, Ferrarese et al. 
2000), with all the possible biases well characterized and quantified (Madore
\& Freedman 1995, hereafter MF95). 
The key observable is the magnitude of the LF sharp cut-off in the 
Cousins I passband, usually
identified by applying an edge-detection algorithm to the LF of the upper 
RGB (see Sakai, Madore, \& Freedman 1996, hereafter SMF96, for a standard 
application).    

As apparent from the CMDs shown in the previous sections, our data allow
for a good sampling of the RGB. In particular, the WF cameras sample 
exclusively the outer halo of the galaxy, a relatively uncrowded region
dominated by low-mass RGB stars and negligibly contaminated by other sources.  
Thus there are favorable conditions to reliably measure I$_{TRGB}$ 
and obtain a distance estimate for NGC~1705
significantly more accurate than the existing ones (O'Connell et al. 1995; 
MFDC; Meurer et al. 1998). 
 
Our WF sample excellently fulfills all the requirements for a safe TRGB 
detection identified by MF95 by means of numerical simulations. 
In particular (1) the number of stars in the upper one mag bin from 
the TRGB is much larger than the minimum indicated by MF95 (100 stars) 
in any of the WF CMDs, (2) the contamination from non-RGB stars is 
negligible,  since in that color range no more than 20 foreground stars and
background galaxies and no member star in other evolutionary phases
can be expected,
and (3) the completeness is much larger than $75 \%$ (indeed, it
is almost $100 \%$) and the effect of blending is negligible
at the position of the tip in the WF cameras (Regions 0-4, see Section 4.2).
The severe crowding and high degree of contamination from brighter 
stars prevent instead a safe identification of the TRGB in the PC data.

\subsection{TRGB detection}

First of all, we have calibrated our data also into the Johnson-Cousins
photometric system following the Holtzman et al. (1995b) prescriptions. 
Then we have applied the 
edge-detection algorithm based on the Sobel's filter to the LF of the upper RGB, 
as in SMF96. The measure has been performed separately for each camera 
and the results are presented in Fig.\ref{figX1}. 
The left panels of Fig.\ref{figX1} report the $I$ vs $V-I$ CMDs 
of the upper RGB for each WF camera
(from top to bottom: WF2, WF3, and WF4, respectively). In these CMDs 
the RGB stands out very clearly and the tip discontinuity is 
evident. The right panels of Fig.~\ref{figX1} show the 
corresponding LFs
(here presented as generalized histograms, see Ikuta \& Arimoto 2000, and
references therein) and the response of the edge-detection filter to these 
LFs.  The TRGB is clearly and uniquely detected in each of the WFs as the 
highest peak in the Sobel Filter Response. The derived estimates
are reported in the plots showing the Sobel Filter Response as a function of the
$I$ magnitude, the quoted errors being the HWHM of the peaks corresponding to 
the TRGB detections. 
The agreement
between the three independent measures is excellent, and we adopt their mean 
as the final estimate, i.e. $I(TRGB) = 24.62 \pm 0.08$. Adopting $E(B-V)=0.045$
from O'Connell et al. (1995) and $A_I\,=\,2.0\,E(B-V)$, derived from 
Dean, Warren, \& Cousins (1978), we finally
obtain $I_0(TRGB) = 24.53 \pm 0.10$, including also a $\pm\, 0.02$ uncertainty
associated to the reddening estimate. 

\subsection{Distance modulus from I(TRGB)}

A serious source of uncertainty for distance moduli estimated via the
TRGB method resides in the calibration of the absolute magnitude $M_I(TRGB)$
as a function of metallicity (see Salaris \& Cassisi 1998 for a 
detailed discussion).
Bellazzini, Ferraro, \& Pancino (2001, hereafter BFP01) recently provided a new 
empirical calibration of $M_I(TRGB)$ as a function of $[Fe/H]$, in the
range $-2.3\le [Fe/H]\le -0.2$. The new calibration is in agreement with
previous ones (Da Costa \& Armandroff 1990; Ferrarese et al. 2000) within the 
uncertainties, but has a much safer and robust observational basis (see BFP01
for details). 

Thus we adopt their eq. 4, averaging over the whole $[Fe/H]$ 
range. In the ``standard'' technique (SMF96) 
the distance modulus and the mean metallicity of the population are
simultaneously estimated by an iterative process. We prefer a more conservative
approach, considering the metallicity as an unknown factor, therefore a mere
source of uncertainty. 
The complete error budget of the final distance modulus is the following: 
$\pm\,0.1$\,mag from the uncertainty in the dereddened
$I_0(TRGB)$, $\pm\,0.06$\,mag due to 
the lack of knowledge of the metallicity of the population, 
$\pm\,0.12$\,mag from the uncertainty on the TRGB calibration (see BFP01), and 
$\pm\,0.2$\,mag as a conservative assumption of the global uncertainty of the
absolute photometric calibration (including uncertainties in zero points,
aperture corrections, short vs. long anomaly, etc.). 

Considering all these possible sources of error, our final best estimate of the 
distance modulus of NGC~1705 is $(m-M)_0\,=\,28.54\,\pm\,0.26$, which 
corresponds to a distance $D\,=\,5.1\,\pm\,0.6$\,Mpc.

At this distance, 1$^{\prime\prime}$ corresponds to 24.8 pc and the
horizontal bar in the bottom right corner of Fig.~\ref{pc_nicmos_bw2}
indicates the apparent length of 100 pc. The PC covers an area of 
893\,$\times$\,893\,pc$^2$ and NIC2 an area of 472\,$\times$\,472\,pc$^2$.

\section{Comparison with stellar evolution models}

For an easier interpretation of the CMD in terms of stellar evolutionary 
phases, in Fig.~\ref{cmdtrackwfpc} we have superimposed on the empirical
optical CMD the Padua stellar evolution tracks with metallicity  
likely to be appropriate for NGC~1705 (Z=0.008, Z=0.004 or Z=0.0004, 
Fagotto et al. 1994a, b). The CMD refers to the 17842 stars of the whole
WFPC2 field, most tightly selected on the basis of DAOPHOT parameters
(i.e. with $\sigma_{DAO}\leq\,$0.1 in both F555W and F814W, and the 
values of $\chi^2$ and {\it sharpness} described in Section~2).
The theoretical tracks have been transformed into the observational plane 
by adopting E(B--V)\,=\,0.045 and (m--M)$_0$\,=\,28.54 (see previous section)
and the tables for bolometric correction and temperature--color 
conversion in the HST VEGAMAG photometric system from Origlia \& 
Leitherer (2000).

The plotted tracks are for masses in the range 0.9 -- 30 \MSUN. Due to
their long lifetimes (in these sets a 0.8 \MSUN~ reaches the TRGB in 19 Gyr),
at the distance of NGC~1705 lower-mass stars wouldn't 
have had time to reach visible phases within a Hubble time. 
For high- and intermediate-mass 
stars we have plotted all the evolutionary phases, while for low-mass 
stars (i.e. $\leq\,$1.7 \MSUN) we have drawn them only up to the TRGB,
to avoid excessive confusion. The MS corresponds to the 
almost vertical 
lines at --\,0.3$\,\leq\,$\mvi$\,\leq\,$--\,0.1 and the turnoff is recognizable 
as a small blue hook. Of the later evolutionary phases, we can clearly
distinguish the almost horizontal blue loops corresponding to core helium 
burning, the bright red sequences of the AGB
of intermediate-mass stars and the fainter red sequences of the RGB of
low-mass stars, terminating at the TRGB with approximately the same luminosity.
The only observed feature which has not a theoretical counterpart is the
almost horizontal tail extending redwards of \mvi\,=\,2 at \mi ~just fainter
than 24. Given its position at the edge of the AGB and TRGB, we tend to
ascribe it to thermally pulsing AGB stars, a poorly understood evolutionary
phase whose models are not displayed in the adopted tracks.

Whatever the assumed metallicity, from the tracks overlap on the empirical
CMD of the whole observed region, it turns out that
the blue plume of NGC~1705 is populated by both intermediate- and high-mass 
stars on the MS and by massive stars at the blue edge of the core helium 
burning phases, the red plume is populated by RSGs and AGB stars, 
and the faint red clump with 1 $\lsim$ \mvi $\lsim $2 and 
24.5 $\lsim$ \mi $\lsim $26.5 is populated by RGB stars. 
It is also apparent that very few stars more massive than 30 \MSUN
~are likely to be present in the resolved field of the galaxy. The more
massive objects present in significant numbers
in the empirical CMD appear to be stars
of 15--30 \MSUN. In fact the theoretical tracks corresponding to 
this mass range can account for several objects spanning in the CMD 
from blue to red colors and connecting the blue and the red plume with
a sort of bright bridge. This suggests that an enhancement 
of the SF activity has probably occurred at the epoch corresponding to 
the lifetime of a 20 \MSUN, i.e. around 15 Myr ago. 

Fig.~\ref{cmdtracknic} shows the analogous superposition of the same three
sets of evolutionary tracks on the most tightly selected CMD derived from
the NIC2 data and containing 829 stars. The strong differential effect 
due to the severe incompleteness
affecting blue faint objects in these bands (see also Al01) results in the
extreme paucity of blue plume stars already mentioned in Section 4.
The red vertical {\it finger} is populated by RSGs and AGB stars, 
while the RGB is too faint to be reachable with these stars so restrictively
selected; we do reach it if we relax the $\sigma_{DAO}$ criterion (see
Fig.~\ref{cm_vjh}) to $\sigma_{DAO}\le 0.2$.
In agreement with what we find for the optical CMD, stellar models
with 15--30 \MSUN ~fit quite well the bridge of brightest points connecting 
the blue and the red plume.

An accurate estimate of the oxygen abundance in NGC~1705 was given by
Storchi-Bergmann et al. (1994), who derive 12\,+\,log\,(O/H)\,=\,8.36 from 
HII regions.
This implies that the youngest objects have a metallicity of about Z\,=\,0.004. 
No direct information is available for older objects. 
The stellar metallicities (Z slightly higher 
or slightly lower than that of the LMC) quoted by MFDC and Sternberg (1998) 
are only indicative.  Wide band photometry is certainly not the best 
way to infer metal abundances, but we can provide some hints.
By simply comparing the color distributions of theoretical tracks and
observed objects in the CMDs of both Fig.~\ref{cmdtrackwfpc} and
Fig.~\ref{cmdtracknic}, it is apparent that the vast
majority of the stars resolved in NGC~1705 are more metal-rich 
than Z\,=\,0.0004, including the old stars on the RGB. In fact, even accounting
for the large uncertainties in the photometric conversions from the
theoretical to the observational plane and for the photometric errors of 
the data, this set of stellar models appears to be too blue to be 
reconcilable with the observed optical and NIR colors in all the red
evolutionary phases. On the other hand,
the Z\,=\,0.008 models seem slightly too red, specially in the red CMD
region. In this case, we cannot exclude that the effect is due
to conversion errors, rather than to a real metallicity excess;
however, we tend to exclude that metallicities higher than this could
be appropriate for this galaxy.

\section{Discussion and conclusions}

We have acquired deep HST photometry of the nearby BCD NGC~1705 to resolve
its stellar populations and derive information on its evolution. With the images 
in F555W, F814W, F110W, and F160W obtained with the use of WFPC2 and NICMOS, we 
are able to detect
for the first time about 40,000 field objects and to sample both the young 
and the old resolved stellar populations. We have also detected several
candidate star clusters (besides the SSC) and background galaxies (including 
a yet unclassified galaxy cluster). These composite objects will be
described in a forthcoming paper (Monelli et al. in preparation), where
also the F380W and F439W images will be analyzed.

The excellent performance of HST allows us to measure also
the fainter/older stars in the RGB phase and to clearly identify in the WFs
the TRGB at \mi$\,\simeq\,$24.5, a 
luminosity level more than 2 mag brighter than our 
most conservative limit, where both incompleteness and blend are not yet
significant. Thanks to this result, we have derived from the TRGB $I$ luminosity
in the Johnson--Cousins photometric system
a distance modulus $(m-M)_0\,=\,28.54\,\pm0.26$ (i.e. a distance of
$5.1\,\pm0.6$ Mpc) in excellent agreement with the modulus 
(28.5$\,\pm\,$0.7) of O'Connell et al. (1994)
and consistent with MFDC's distance (4.7 Mpc).

NGC~1705 is considered a {\it post-starburst} galaxy, 
because even the SSC, where the SF activity has been more concentrated and 
recent, has stopped forming stars at least 5--6 Myr ago, as argued by Heckman
\& Leitherer from the lack of spectral features from O and Wolf--Rayet 
(WR) stars.
Our data show that also outside the SSC there are only a few stars 
more massive than 30 \MSUN, and most of them seem to be connected to
the brightest SSC, through the pearl necklace shown in Fig.~\ref{pc_nicmos_bw2}. 
This paucity of very massive stars is in agreement with MFDC who identified only
5 candidate HII regions and one extended ionizing association ({\it g+h}
in their nomenclature) with emission-line features indicating the presence
of WR stars. We thus suggest that the most recent conspicuous SF activity 
in the field of NGC~1705 has occurred approximately 10--20 Myr ago. 

NGC~1705 definitely contains a a significant population of stars several
Gyr old.
This does not imply, though, that the galaxy population is essentially
old. By separating, with the help of the Padua stellar 
evolution tracks, the observed optical CMD in three different zones
corresponding to different stellar masses, we find that in the PC field, 
which at our derived distance covers
a region of about 893$\,\times\,$893\,pc$^2$, of the 9224 stars 
most tightly selected, 50\% have masses $\geq\,$3 \MSUN ~and, hence, age lower
than 500 Myr, and $\sim\,$800 (9\%) have masses $\geq\,$7 \MSUN, and are 
thus younger than 60 Myr. 

From {\it bona fide} RGB stars, we can estimate also the fraction of the 
oldest objects. As apparent from Fig.~\ref{cmdtrackwfpc}, the color
of the RGB depends strongly on the metallicity: if Z\,=\,0.008, the blue edge
of our resolved RGB stars is \mvi$\,\simeq\,$1.4, if Z\,=\,0.004, it 
is around 1.2,
and, if Z\,=\,0.0004, it is around 1.0. If we consider as RGB stars all those
with \mi$\,\geq\,$24.5 and \mvi ~redder than these blue edges, we find that
3511, 2398 or 1227 (for Z\,=\,0.008, 0.004, or 0.0004, respectively) 
over 9224 stars are older than 1 Gyr. In other words, thanks to the higher
resolving power of the HST, we have been able to measure also in the inner 
893$\,\times\,$893\,pc$^2$ the oldest population, which was too faint to be
resolvable with ground-based observations.

The diagrams of Fig.~\ref{cm_vi} confirm the existence of an age gradient.
We find that:
\begin{itemize}
\item  the most massive stars (M$\lsim$30 \MSUN), younger than 10--20 Myr,
are all located within 100 pc from the galactic center and the most luminous 
SSC, most of them in Region 7 and a few in Region 6;
\item intermediate-mass stars with ages up to 1 Gyr are visible 
only in Regions 6 and 5, i.e. within $\sim\,$500 pc from the center. Some of
them are present also in Region 7, but the extreme crowding and high background
of this zone makes their detection highly improbable;
\item beyond $\sim\,$500 pc from the center, the galaxy is populated only by
low mass stars on the RGB, i.e. with ages from a few Gyr to a Hubble time.
\end{itemize}
As mentioned in the Introduction, previous studies already pointed
out that NGC~1705 has a composite population (Quillen et al. 1995), with
an age gradient (MFDC) like the one here described. This is, however, the
first evidence from direct analysis of the resolved stellar populations.
The same kind of spatial segregation of the younger stars has been found also in
other late-type dwarfs (e.g., NGC~1569, G98; I Zw 18, ATG; VII
Zw 403, Schulte-Ladbeck et al. 1999a).

In summary, NGC~1705 is definitely not a starburst galaxy, since it does
not appear to have had any conspicuous SF activity in the last few Myrs,
neither in the field nor in the SSC. Our CMDs show that the last episode
of significant SF in the field has occurred around 
10--20 Myr ago, preceded by several
other episodes or by a continuous activity. To understand whether or not 
any of these episodes can be considered as a real burst (i.e. short and 
intense SF activity), and how long the quiescent phases (if any) have lasted,
one must perform a more sophisticated analysis. We are working on this
analysis using the method of synthetic CMDs described by Tosi et al. (1991)
and G98, and the results will be presented in a forthcoming paper (Annibali
et al. in preparation). 

The impressive morphology of the H$_{\alpha}$ images of the galaxy,
dominated by loops and arcs apparently centered on the brightest SSC, and the
corresponding kinematics have been studied by MFDC. They interpreted the 
loops and arcs as hot expanding bubbles energized by supernova ejecta 
and stellar winds from its nucleus, with estimated expansion timescale of
$\sim$9 Myr. Recent FUSE observations of NGC~1705 (Heckman et al. 2001) 
also show that the superbubbles will most probably blow out of the galaxy
and remove a significant fraction of its metals.
Sahu \& Blades (1997) agreed with this scenario of conspicuous
galactic wind triggered by supernova explosions, on the basis of
HST UV spectra. Their data confirmed the 540 km s$^{-1}$ velocity 
of the supershell
and showed the overabundance in the galaxy gas of elements produced by massive
stars. X-ray observations (Hensler et al. 1998) supported the same picture,
and the discovery of an extremely soft component in the X-ray emission led 
these authors to suggest that the SSC is not located at the galactic
center, a possibility already
considered by Heckman \& Leitherer (1997) and confirmed by our images
and by the isophotal contours shown in Fig.~\ref{mosaico_bw}. It would
be interesting to check whether the pearl necklace of bright young stars
and the asymmetric distribution of the candidate star clusters displayed
in Fig.~\ref{pc_nicmos_bw2} are physically related to the SSC impact on
the surrounding environment.

Quantitative information on the star formation history of NGC~1705 will be 
useful also to assess the effect 
of galactic winds on the chemical enrichment of its ISM. This galaxy 
has clearly formed stars over several Gyrs, most of its old stars have a 
fairly high metallicity of Z\,=\,0.004, and yet it is characterized
by strong winds currently removing heavy elements from its ISM. Was the
early metallicity high because of a very strong initial SF ?
Is the current wind event the first one occurring over the galaxy lifetime ? 
If this is the case, why have not early SF episodes been able to trigger the 
winds (especially if the earliest activity was strong) ?
Or does it mean that winds do not significantly affect
the galaxy metallicity (for instance, because they involve too small amounts
of gas, or because after some time the metals fall back on the galaxy) ?
To try to answer these questions, quantitative estimates of the SF rate and
of the IMF are necessary. Our forthcoming results from the synthetic CMD
method will provide such information, since we will 
derive the epoch, the duration and the intensity of each major SF episode, 
and we will infer indications on the most likely IMF. 

Very preliminary results (Annibali et al. 2001) indicate that indeed 
NGC~1705 has been forming stars since 10--15 Gyr, with different rates at 
different epochs, but without evidence of strong bursts or really
quiescent intervals. They also suggest that the IMF
of the young stellar populations is flatter than normal (with a slope 
$\alpha\,\simeq\,$1.5 in the usual approximation $dN/dm\,\propto\,m^{-\alpha}$, 
where Salpeter's slope is 2.35), therefore implying a fraction of massive 
stars higher than normal. This unusual IMF, even if relative to the field 
stars and not to the SSC stellar population, 
is in agreement with one of the two 
Sternberg's (1998) arguments (the second one being a truncated 
IMF at a lower mass 
limit between 1 and 3 \MSUN) to explain the SSC overluminosity, 
which at our derived distance of 5.1 Mpc is indeed confirmed. More accurate and 
extensive tests in the various Regions, as well as the photometric reduction 
and analysis of the F380W and F439W data, are however necessary to reach firmer 
conclusions and are currently being performed (Annibali et al. in
preparation; Monelli et al. in preparation).

\acknowledgments

We thank Luis E. Bergeron, Stefano Casertano, Mark Dickinson, Luciana Federici, 
and Matteo Monelli for help and suggestions to overcome the data reduction problems. 
Daniela Calzetti, Mark Clampin, Jay Gallagher, Antonella Nota, Livia Origlia, and
Lucia Pozzetti are warmly thanked for useful information and discussions. This work 
has been partly supported by the Italian ASI, through grants ARS-96-70 and ARS-99-44, 
and by the Italian MURST, through Cofin2000. Support for this work was also provided 
by NASA through grant number GO-07881.01-96A from the Space Telescope Science Institute.
Funding for A.A. has also been provided through an Italian CNAA fellowship. E.S. thanks 
the STScI for the hospitality during a one month visit, partially funded by a fellowship 
from the University of Bologna.

\clearpage

\figcaption{Mosaic of WFPC2 images of NGC~1705 in the F814W filter. North is up
and the axes of the equatorial coordinate system are indicated. The field 
of view of the PC is 36$^{\prime\prime} \times 36^{\prime\prime}$,
while that of each WF is 80$^{\prime\prime} \times 80^{\prime\prime}$.
Isophotal contour levels are superimposed on the image (see text for details).
\label{mosaico_bw}}

\figcaption{Right hand panel: NIC2 image of NGC~1705 in the F110W filter.
Left hand panel: enlargement of a portion of the PC image of 
Fig.~\ref{mosaico_bw}, with superimposed the box of the NIC2 field and
the inner isophotal contours. Open circles identify candidate 
star clusters (see text). The bar in the bottom right corner gives the linear
scale of the image at an assumed distance of 5.1 Mpc.
\label{pc_nicmos_bw2}}

\figcaption{DAOPHOT photometric errors $\sigma_{DAO}$ vs. calibrated 
magnitude in all filters: for optical filters the PC data have been plotted 
separately from the WF ones because of the different S/N.
\label{dao_sig}}

\figcaption{CMDs of the stars measured in WFPC2 fields and selected with 
the criteria described in the text. The stars have been divided into six 
groups on the basis of the contour levels of Fig.~\ref{mosaico_bw}. 
Regions 1, 2 and 3, 4 are grouped together because of their similar
stellar populations and crowding conditions. The number of stars
plotted in each box is labelled in the bottom right corner. The average size
of the photometric errors, as derived from the artificial stars experiments,
are indicated. The color error is evaluated at \mvi=1.
\label{cm_vi}}

\figcaption{CMDs of the stars measured in the NIC2 field. 
The top panels show \mh~vs \mj\,--\,\mh~for all the objects selected in the two 
bands with the criteria described in the text. These have been divided
into three groups on the basis of the contour levels of Fig.~\ref{mosaico_bw}. 
The number of the stars plotted in each box is given in the bottom right 
corner. The average size
of the photometric errors, as derived from the artificial stars experiments,
are indicated. The color error is evaluated at \mjh=1.
The lower panels show \mh~vs \mv\,--\,\mh~for the objects identified in all the
four bands. The color error is evaluated at \mv -- \mh~=2.
\label{cm_vjh}}

\figcaption{Diagrams of magnitude differences (input--output) vs input 
magnitude from the artificial star experiments in the F555W  filter. 
From the outer 
to the innermost Region, the plotted stars are 93260, 18638, 23581, 17477, 
3132 and 108, respectively. The standard deviations in 1 mag bins 
around \mv~= 23, 25, 27 are indicated for each Region. The lines superimposed
on the diagrams represent the local mean $\Delta m$ and the $\pm$ 1 standard
deviations.
\label{dm_v}}

\figcaption{Same as Fig.~\ref{dm_v} for the F814W frame. The number of 
plotted stars is 86770, 16722, 22116, 18912, 5020, 373. The standard deviations
around \mi~=23, 25 and 27 are also given.
\label{dm_i}}

\figcaption{Same as Fig.~\ref{dm_v} for the F110W (left panels) and F160W
frames (right panels). The stars plotted for the F110W filter are
73601, 45388, 3500 
from the bottom to the top panel, and their standard deviations are 
computed at \mj~= 20, 22, 24. The stars in the F160W filter are 47745, 31815, 
2638 and the standard deviations are computed at \mh~=19, 21, 23.
\label{dm_jh}}

\figcaption{Completeness levels superimposed on the CMDs of 
Fig.\ref{cm_vi} for the six WFPC2 Regions. The levels are derived combining
the completeness factor in the \mv~ band with
that of the \mi~ band (see text). 
From top to bottom of each box the plotted lines correspond to completeness
factors $CF=0.95$ (thick line), $CF= 0.75$ (thin line),
$CF= 0.50$ (thick line), and $CF= 0.25$ (thin line). 
An evenly spaced grid of blending vectors is also superimposed to the
CMDs (thin lines with black dots) . The amplitude of these vectors is
the average $\Delta m$ described in the text, the dots indicate the
direction of the vectors. The effect of such vectors become appreciable only for
$m_{F814W}\ge 24$ in the regions 5 and 6, and $m_{F814W}\ge 26$ for regions
0-4 and only at extreme colors.
\label{cm_corr_vi}}

\figcaption{Same as Fig.~\ref{cm_corr_vi} but for the CMDs of 
Fig.\ref{cm_vjh}.
\label{cm_corr_jhv}}

\figcaption{Detection of the TRGB. Left panels: $(I,V-I)$ CMDs of the upper RGB
for the three WF cameras. 
Right panels: the corresponding Luminosity Functions,
shown as generalized histograms, coupled with the response of the Sobel Filter
to the LFs. The peaks in the Sobel Filter response correspond 
to the position of the
TRGB.
\label{figX1}}

\figcaption{Optical CMDs of the 17842 stars selected in the WFPC2 fields
with superimposed the Padua stellar evolutionary tracks 
(Fagotto et al. 1994a, b). The tracks in the top panel have metallicity 
Z=0.008, those in the middle one have Z=0.004 and those in the bottom 
panel have Z=0.0004. Only the following stellar masses are shown: from 
left to right, 30, 15, 9, 7, 5, 4, 3, 2, 1.8, 1.6, 1.4, 1.2, 1.0, 0.9 \MSUN.
The corresponding lifetimes are (they slightly differ from 
one metallicity to the other) 7, 15, 35, 56, 110, 180, 370, 1120, 1300,
1630, 2520, 4260, 8410, and 12500 Myr, respectively.
\label{cmdtrackwfpc}}

\figcaption{Same as Fig.\ref{cmdtrackwfpc}, but for the NIR CMD of the 829
stars selected in the NIC2 field. 
\label{cmdtracknic}}


\plotone{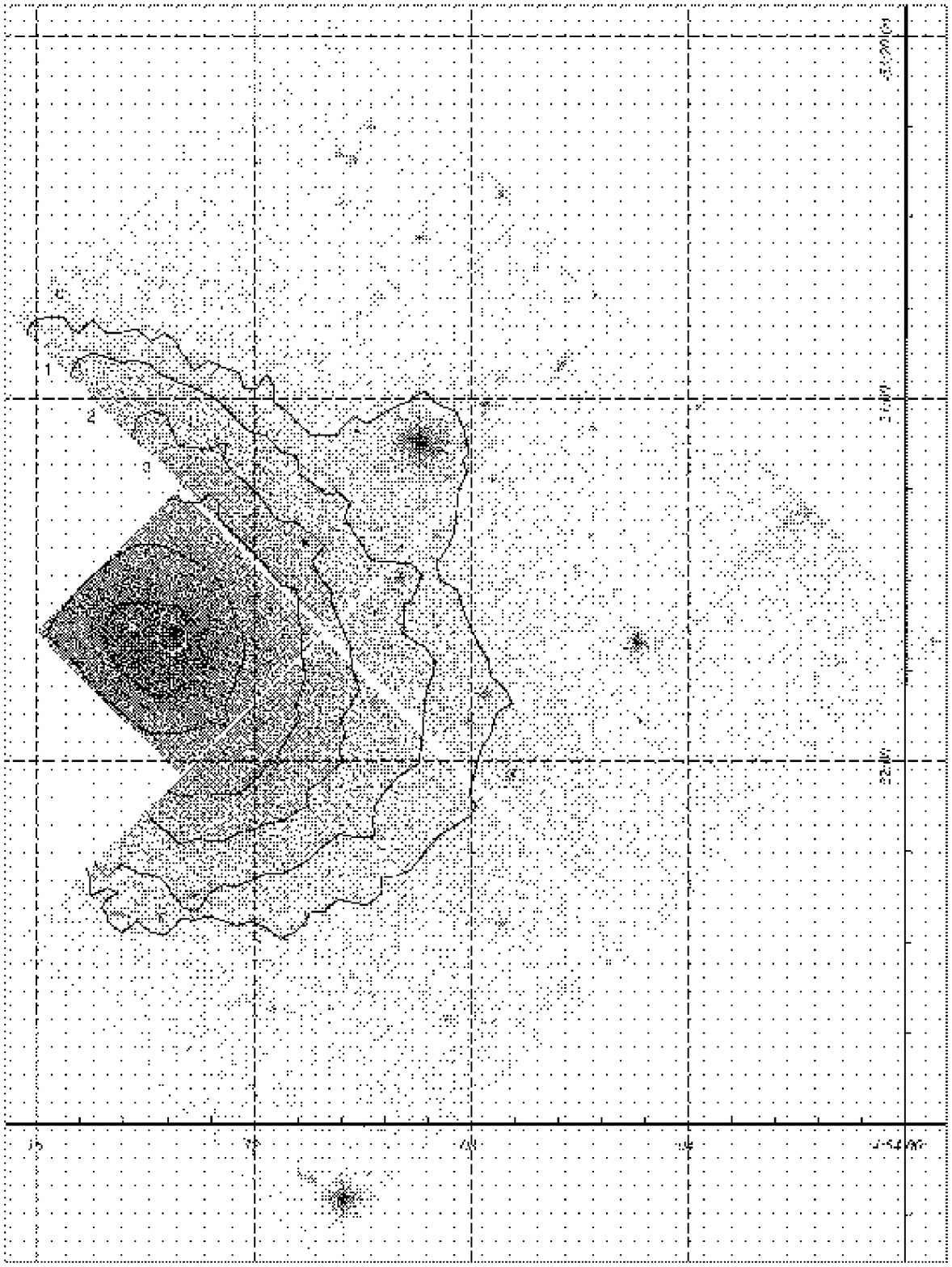}  
\plotone{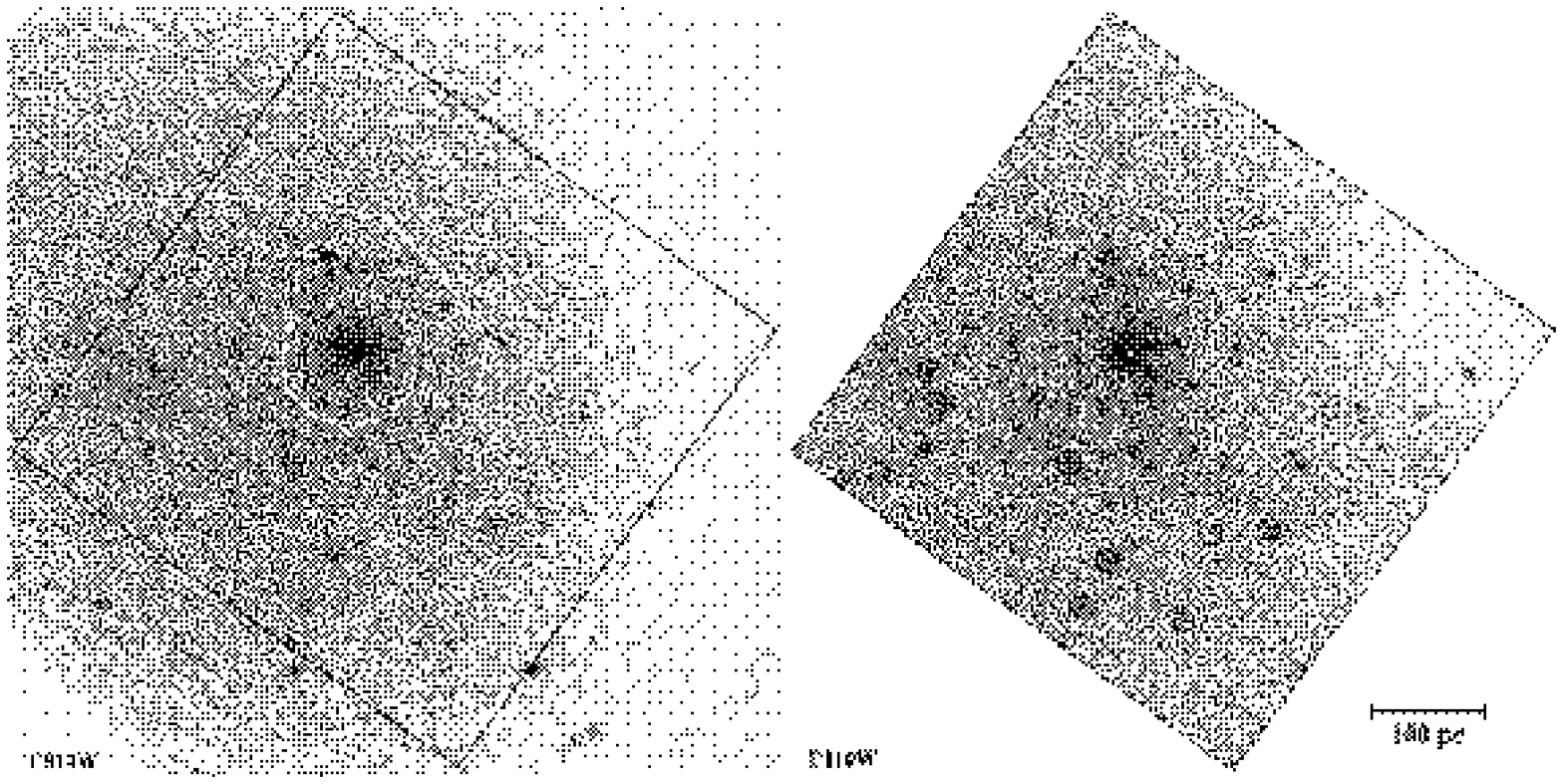}  
\plotone{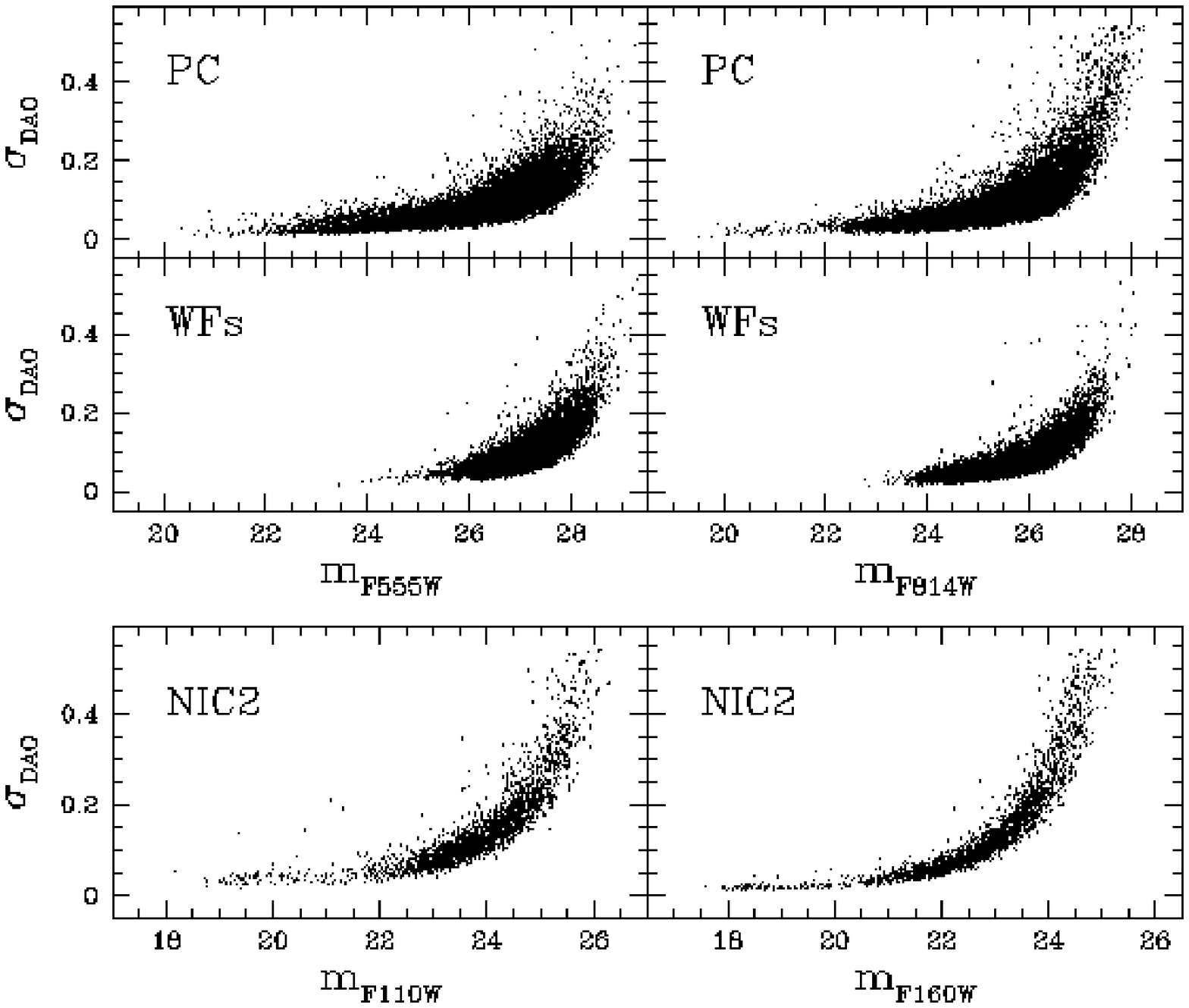}  
\plotone{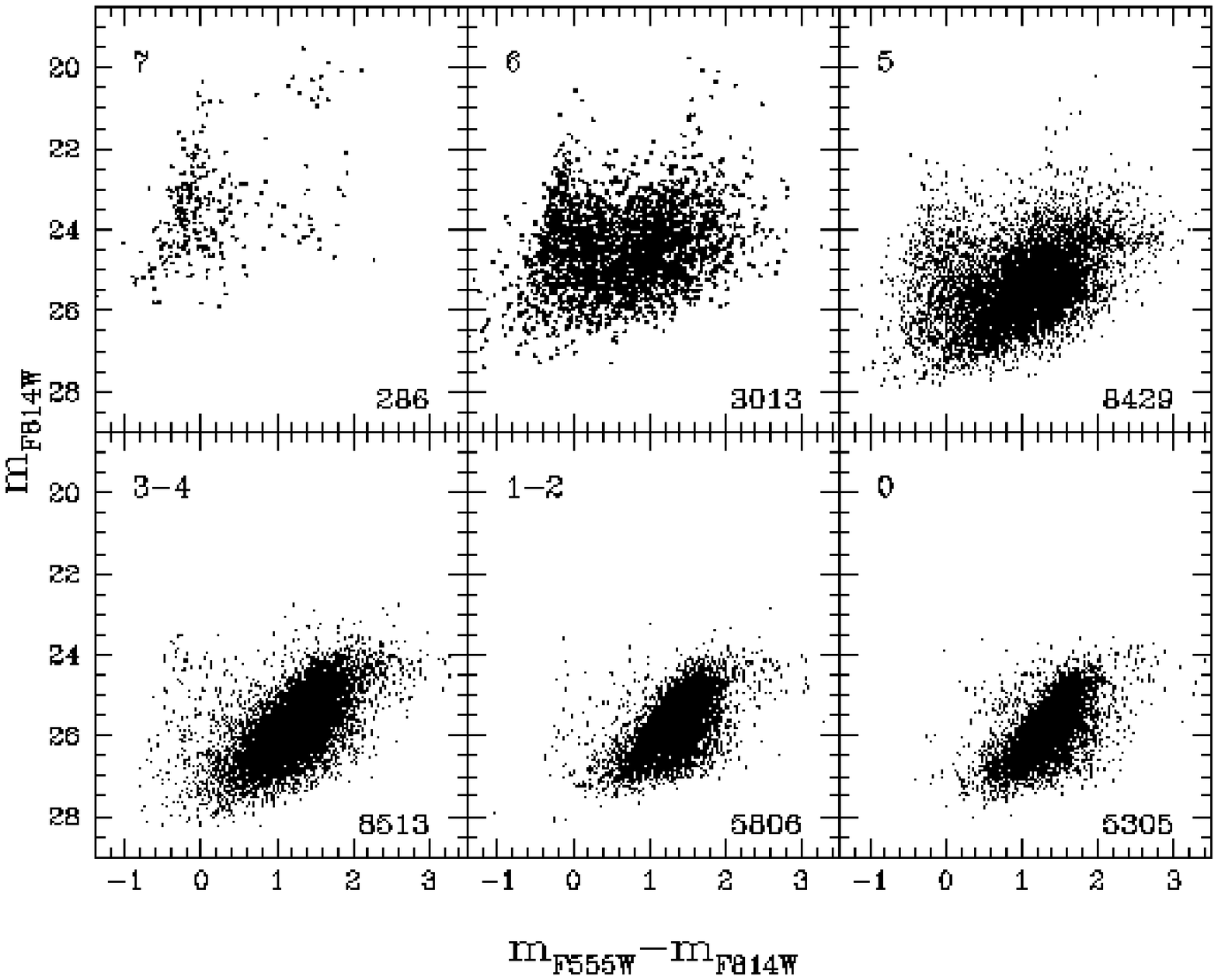}  
\plotone{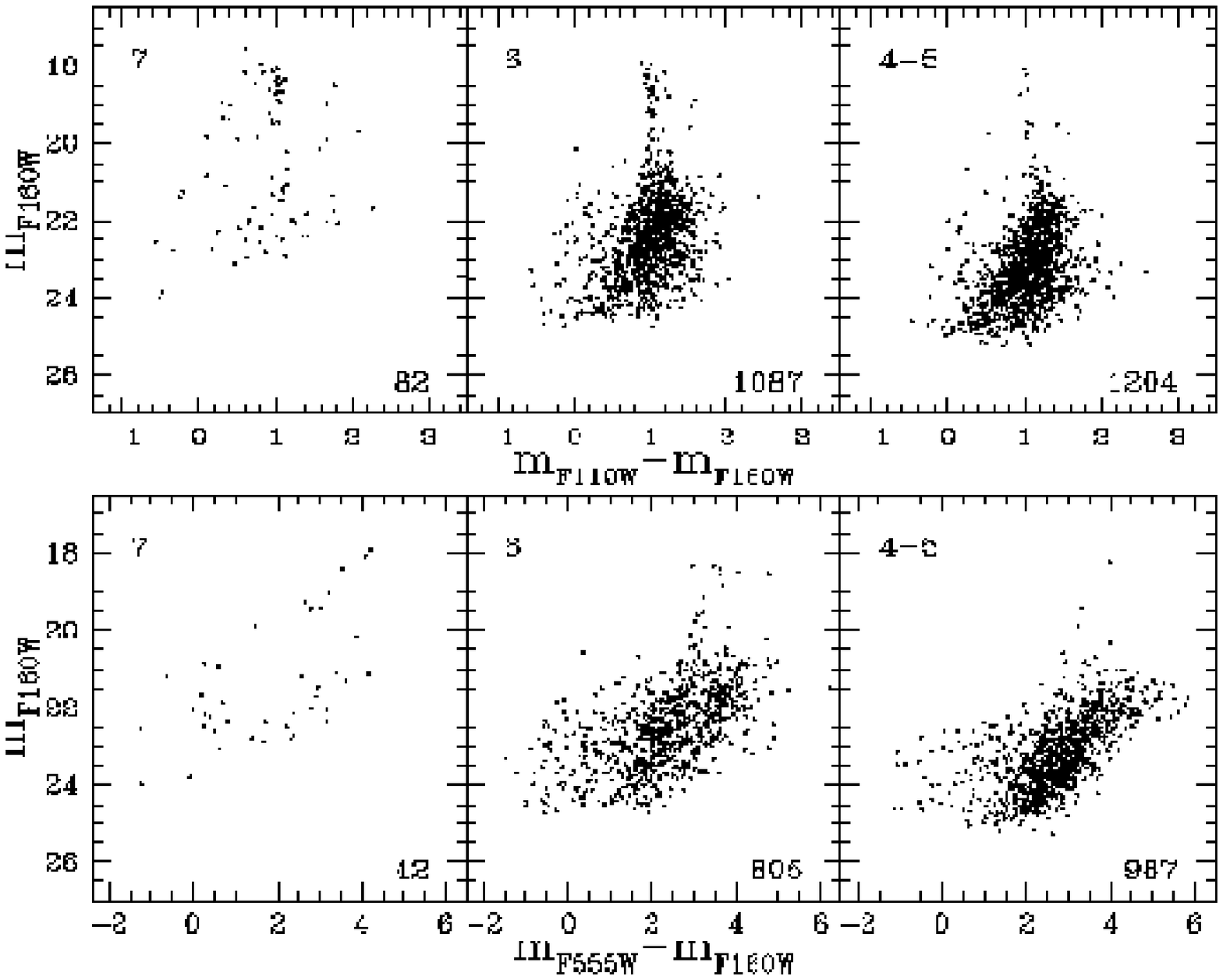}  
\plotone{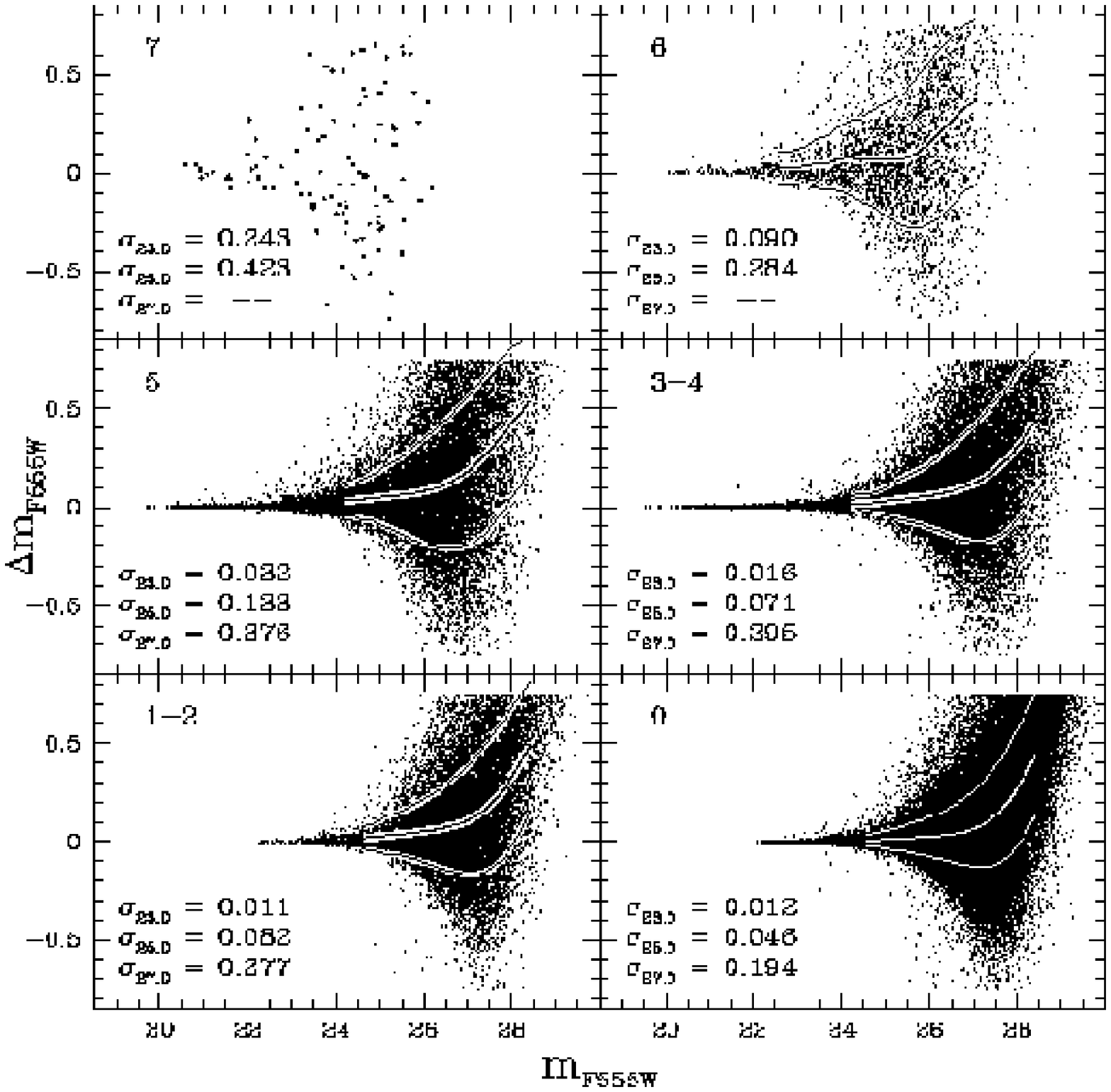}  
\plotone{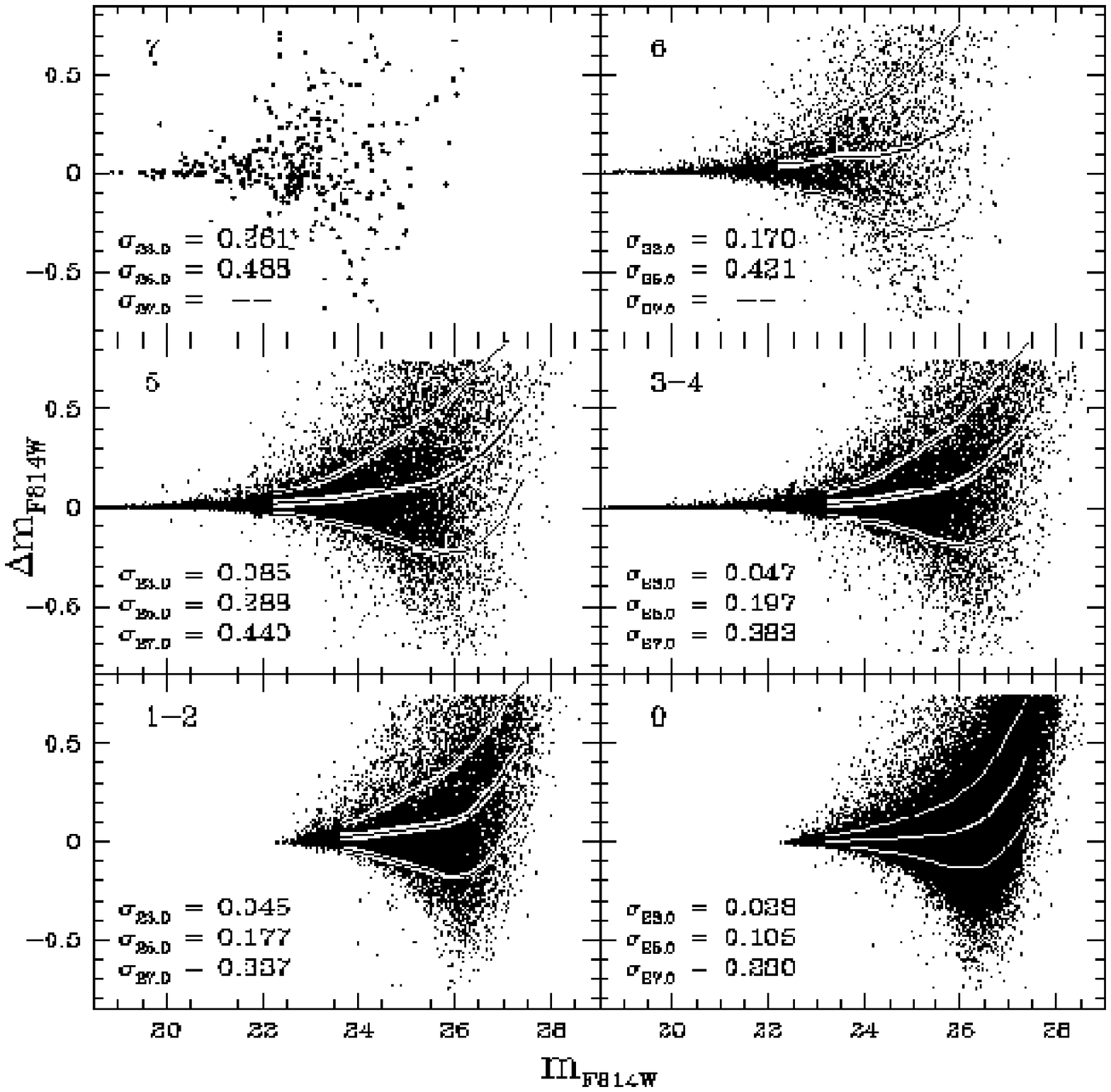}  
\plotone{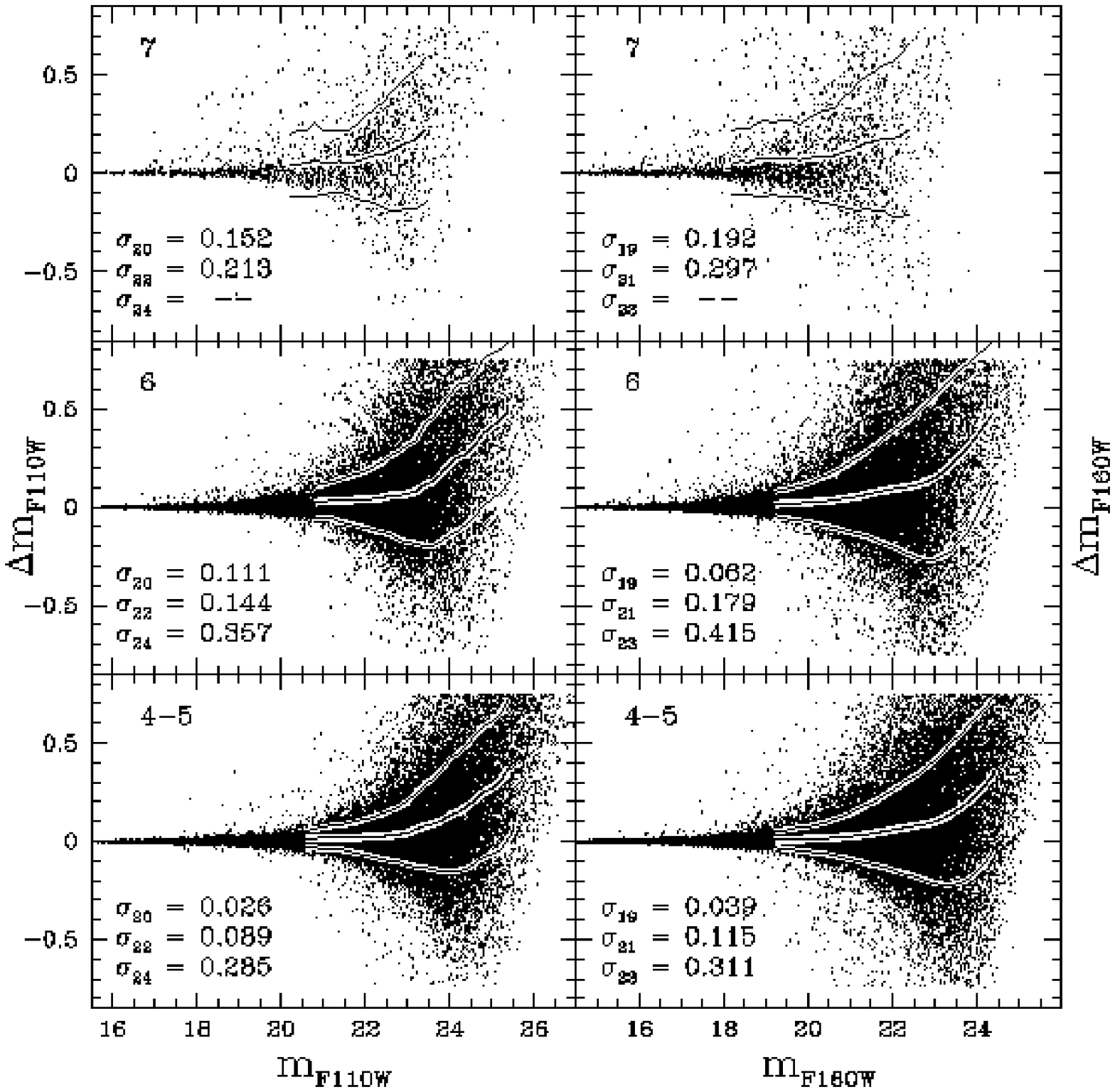}  
\plotone{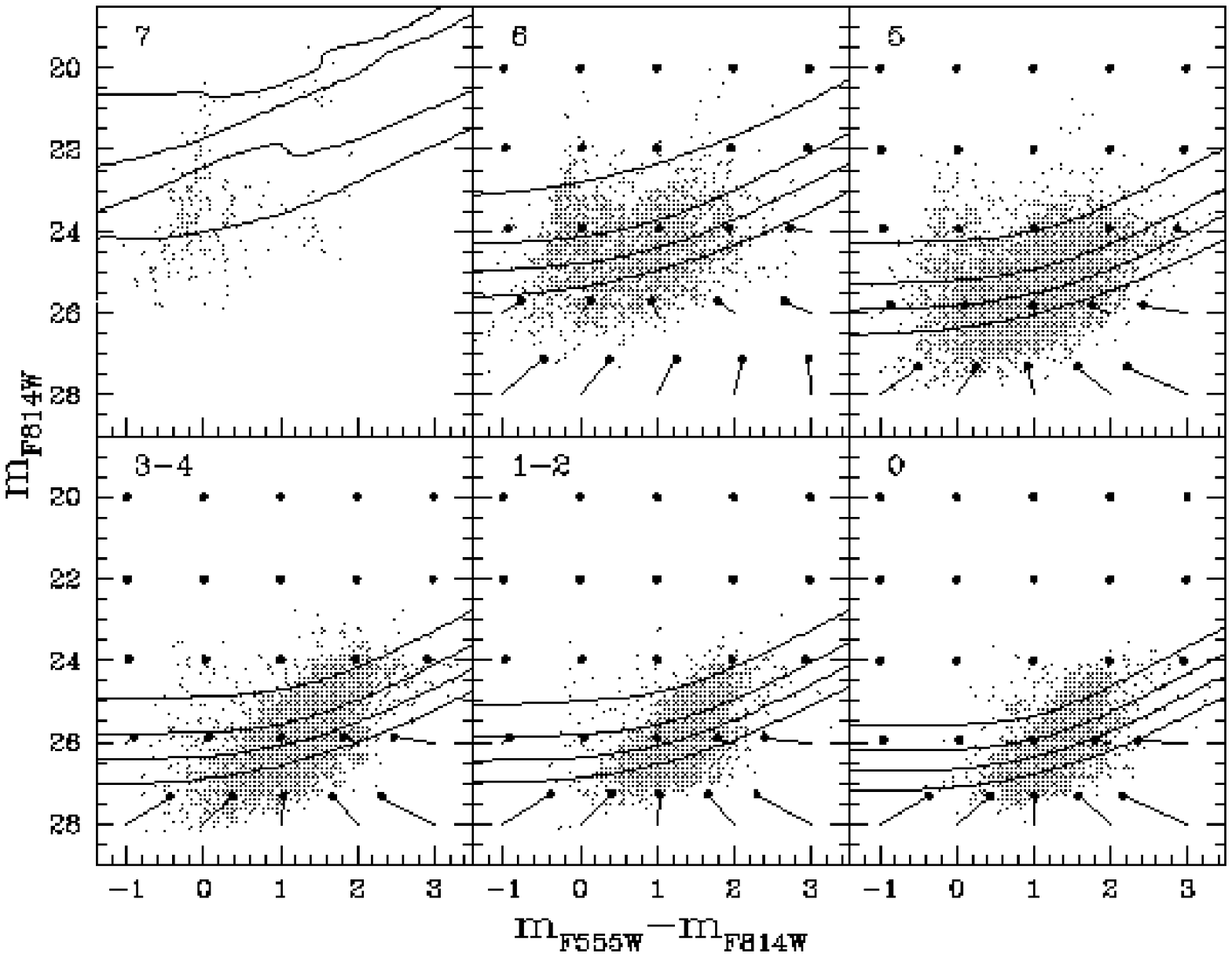}  
\plotone{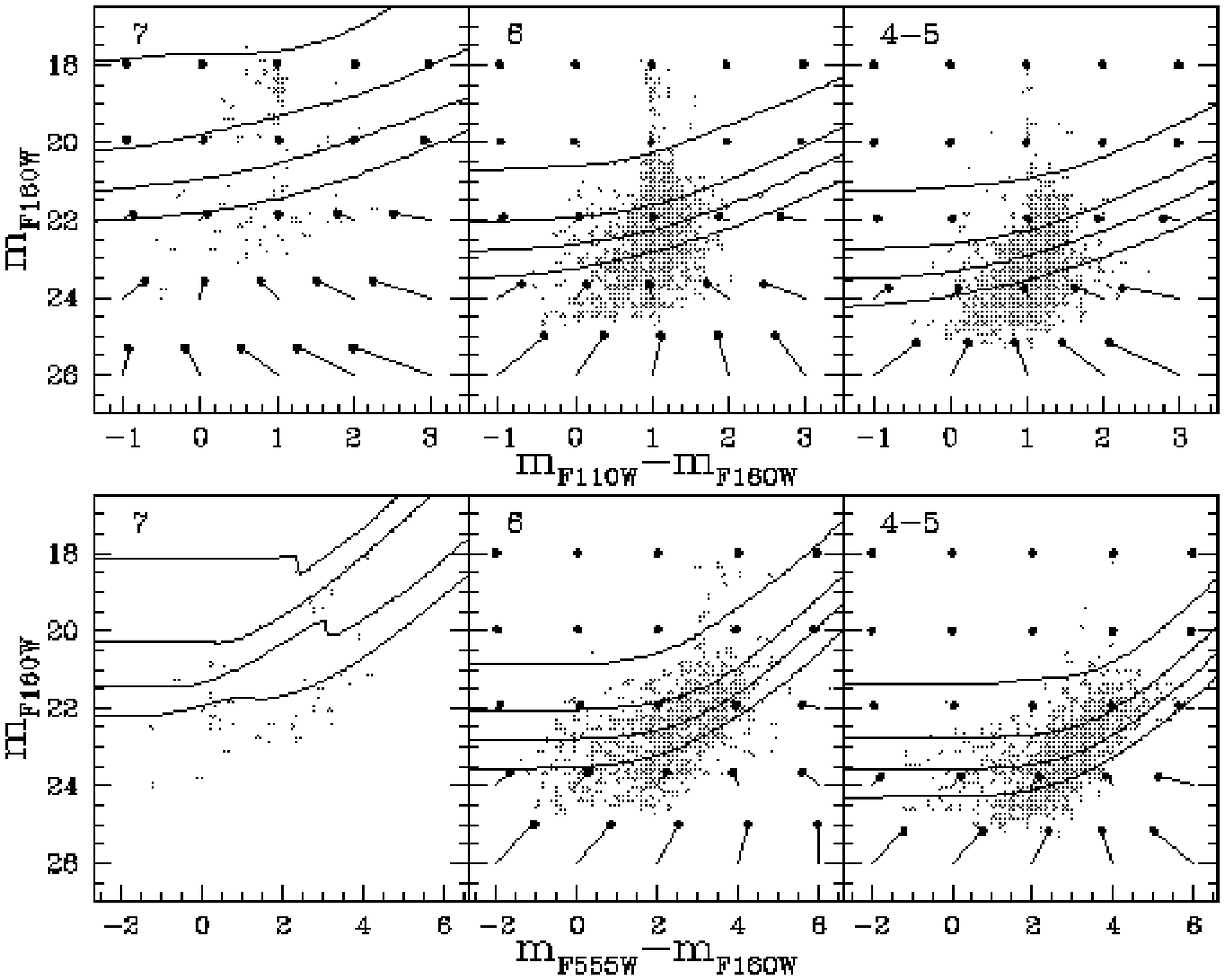}  
\plotone{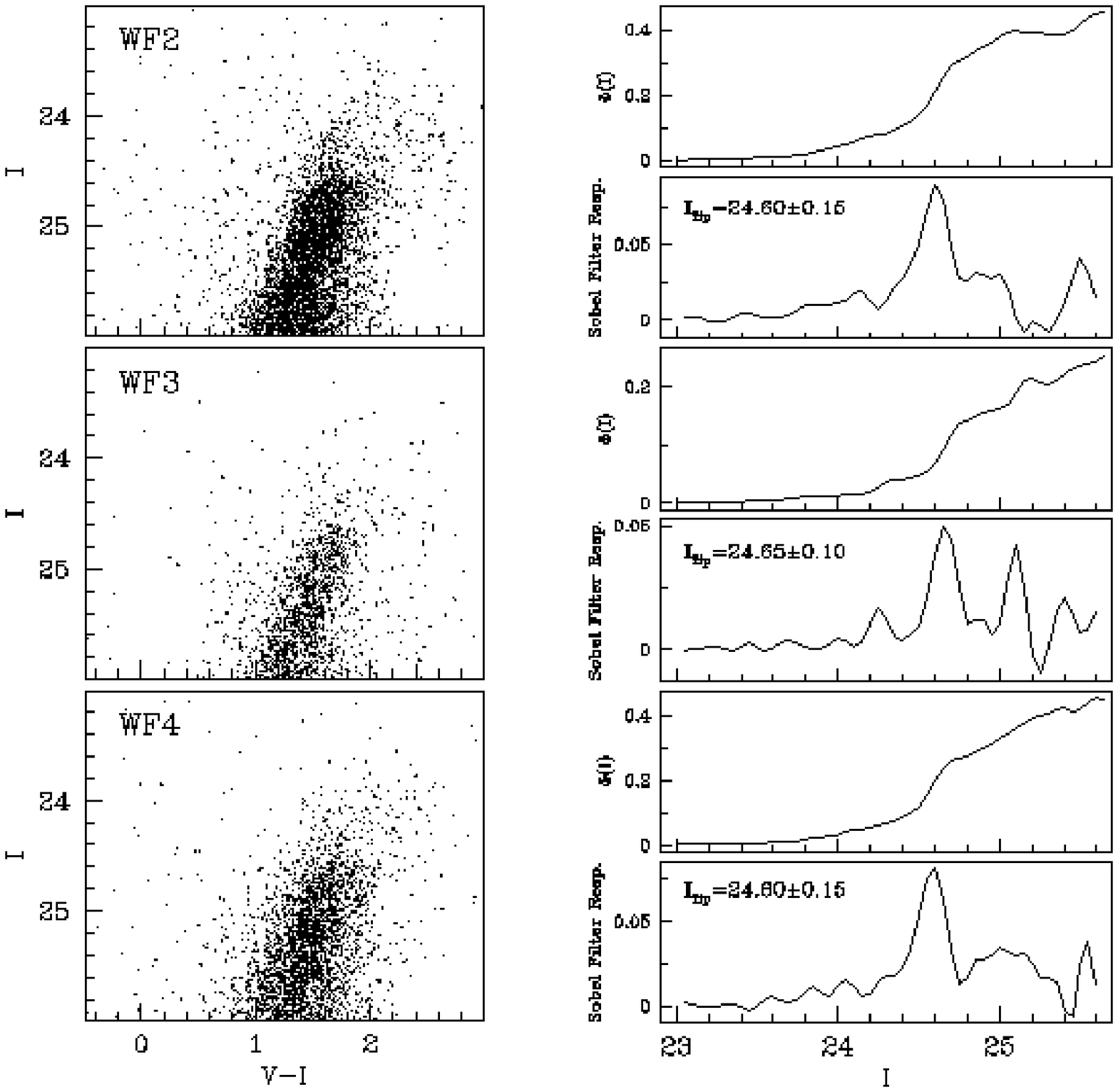}  
\plotone{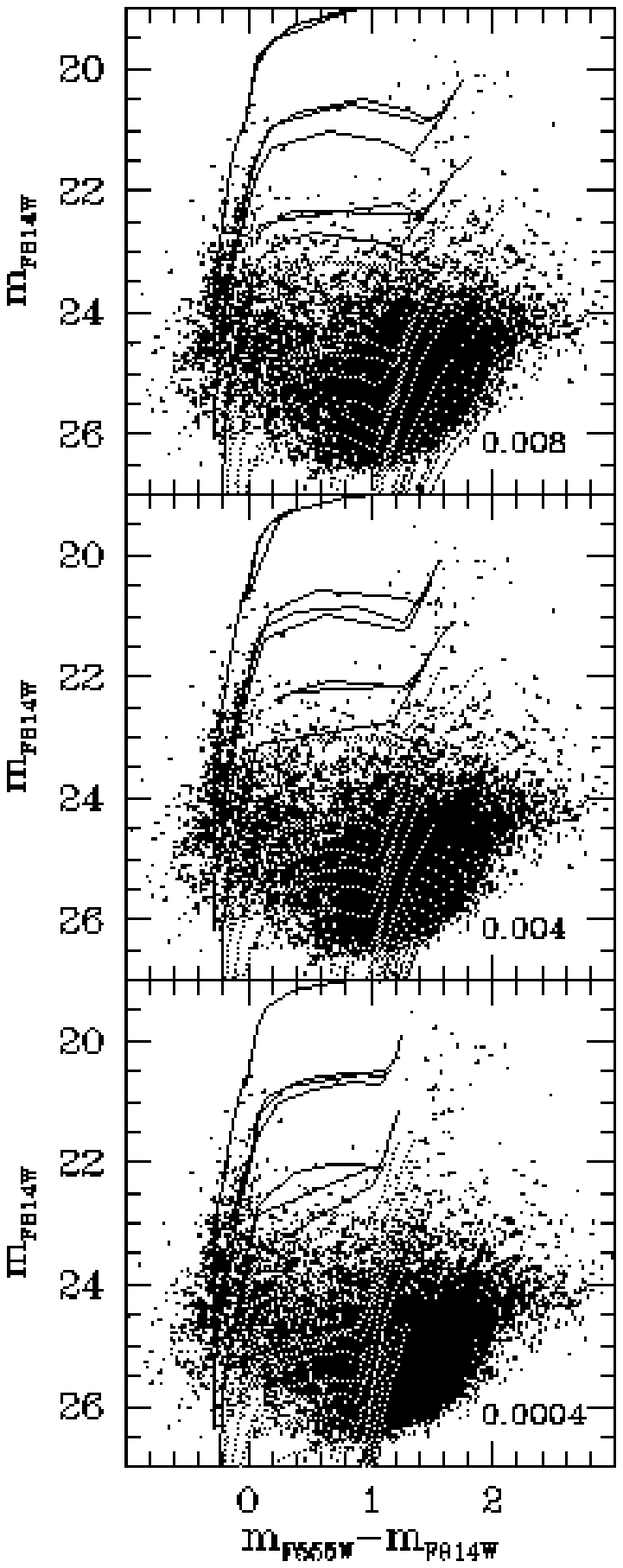}  
\plotone{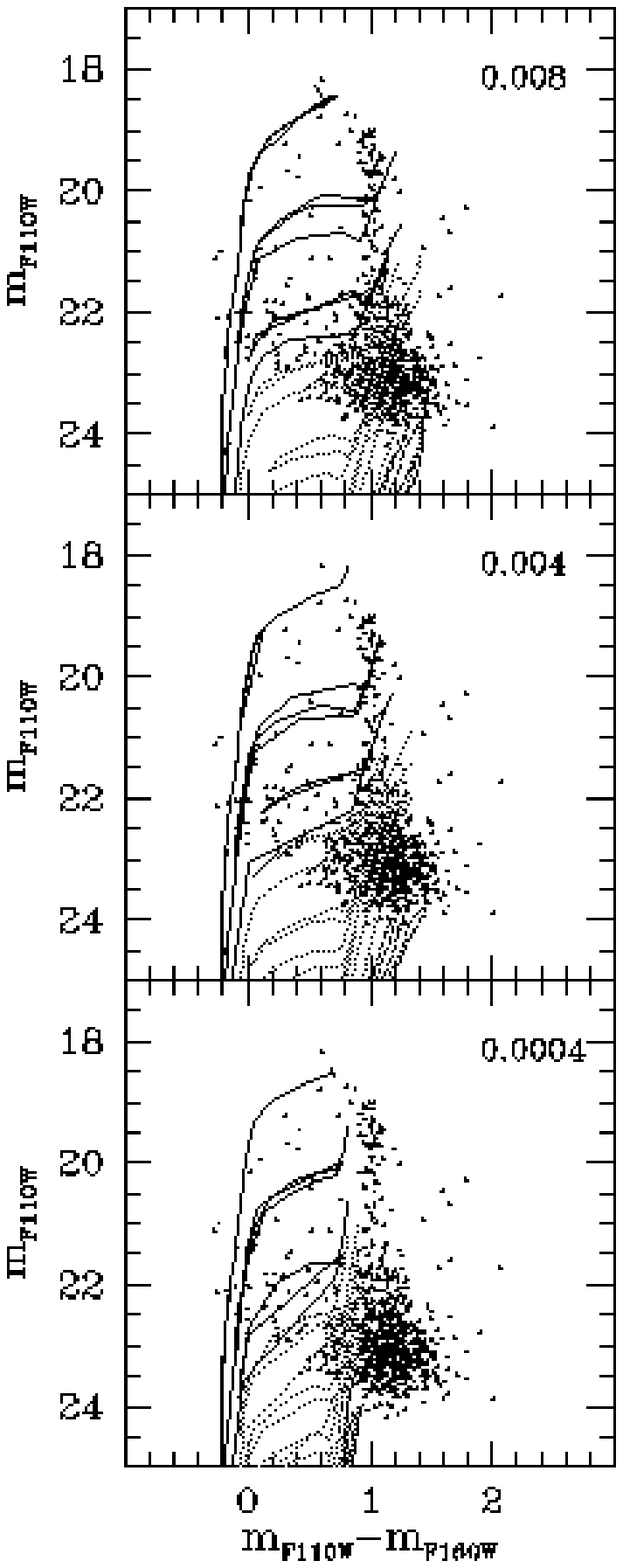}  


\end{document}